\documentclass[a4paper,11pt]{article}
\usepackage{graphicx}
\usepackage{epsfig}
\usepackage{amsmath}
\usepackage{amssymb}
\usepackage{mathtools}

\DeclarePairedDelimiter{\ceil}{\lceil}{\rceil}

\newcommand{\ds}{\displaystyle }

\newcommand{\beq}{\begin{equation}\it }
\newcommand{\eeq}{\end{equation}}

\title{Generalized space-time fractional dynamics in networks and lattices}
\author{ { T.M. Michelitsch$^{1}$\footnote{Corresponding author,
e-mail~: michel@lmm.jussieu.fr },
A.P. Riascos$^2$, B.A. Collet$^{1}$, A.F. Nowakowski$^3$, F.C.G.A. Nicolleau$^3$  }
\\ \\
$^1$ Sorbonne Universit\'e \\ Institut Jean le Rond d'Alembert, CNRS UMR
7190 \\
4 place Jussieu, 75252 Paris cedex 05, France
\\ \\
$^2$ Instituto de F\'isica, Universidad Nacional Aut\'onoma de M\'exico, \\
Apartado Postal 20-364, 01000 Ciudad de M\'exico, M\'exico
\\ \\
$^3$ 
Department of Mechanical Engineering \\
University of Sheffield \\
Mappin Street, 
Sheffield S1 3JD,
United Kingdom 
}

\begin{document}

\maketitle

\begin{abstract}
We analyze generalized space-time fractional motions on undirected networks and lattices. 
The continuous-time random walk (CTRW) approach of Montroll and Weiss is 
employed to subordinate a space fractional walk to a generalization of the time-fractional Poisson renewal process.
This process introduces a non-Markovian walk with long-time memory effects and fat-tailed characteristics 
in the waiting time density.
We analyze `generalized space-time fractional diffusion' in the infinite $\it d$-dimensional 
integer lattice $\it \mathbb{Z}^d$. We obtain in the diffusion limit a `macroscopic' space-time 
fractional diffusion equation.
Classical CTRW models
such as with Laskin's fractional Poisson process and standard Poisson process which occur as special cases are also analyzed.
The developed generalized space-time fractional CTRW model contains a four-dimensional parameter space and offers therefore a great 
flexibility to describe real-world situations in complex systems.
\end{abstract}
\section{\small INTRODUCTION}
\label{intro}

Random walk models are considered to be the most fundamental approaches to describe stochastic processes in nature. 
Hence applications of random walks cover a very wide area in fields as various as random search strategies, 
the proliferation of plant seeds, the spreading phenomena of pandemics or pollution, chemical reactions, finance, 
population dynamics, properties of public transportation networks, anomalous diffusion and generally such approaches are able to capture empirically observed power-law features in `complex systems' 
\cite{Zaslavsky2002,Shlesinger2017,SaichevZaslavski1997,Gorenflo2007,MetzlerKlafter2000,MetzlerKlafter2004,BarkaiMetzlerKlafter2000}. 
On the other hand the emergence of ``network science'', and especially the study of random walks 
on networks has become a major subject
for the description of dynamical properties in complex systems \cite{TMM-APR-ISTE2019,MasudaPorterLambiotte2017,RiascosWangMiMi2019}.

In classical random walks in networks, the so-called `normal random walks', the walker in one step can reach only connected next neighbor sites \cite{Polya1921,NohRieger2004}. 
To the class of classical Markovian walks refer continuous-time random walks (CTRWs) where the walker undertakes jumps 
from one node to another where the waiting time between successive jumps are exponentially distributed leading to Poisson distributed numbers of jumps.
The classical CTRW models with Poisson renewal process are able to capture normal diffusive properties
such as the linear increase of the variance of a diffusing particle \cite{MetzlerKlafter2004}. However, 
these classical walks are unable to describe power-law features exhibited by many complex systems such as the 
sublinear time characteristics 
of the mean-square displacement in anomalous diffusion \cite{MetzlerKlafter2000}. It has been demonstrated that such anomalous 
diffusive behavior is well described by a random walk subordinated to the fractional generalization of the Poisson process. 
This process which was to our knowledge first introduced by Repin and Saichev \cite{RepinSaichev2000}, 
was developed and analyzed by
Laskin who called this process the `fractional Poisson process' \cite{Laskin2003,Laskin2009}. 
The Laskin's fractional Poisson process was further generalized in order to obtain greater flexibility to adopt real-world situations
\cite{PolitoCahoy2013,MichelRiascos2019,MichelitschRiascosGFPP2019}. 
We refer this renewal process to as `generalized fractional Poisson process' (GFPP). 
Recently we developed a CTRW model of a normal random walk subordinated to a GFPP \cite{MichelRiascos2019,MichelitschRiascosGFPP2019}.

The purpose of the present paper is to explore space fractional random walks that are subordinated to a GFPP. We analyze such motions in undirected networks and as a special application
in the multidimensional infinite integer lattice $\it \mathbb{Z}^d$.

\section{\small RENEWAL PROCESS AND CONTINUOUS-TIME RANDOM WALK}
\label{RenewalProc}

In the present section, our aim is to give a brief outline of renewal processes (or also referred to as `compound processes') and closely related to the `continuous-time random walk
(CTRW)' approach which was introduced by Montroll and Weiss \cite{MontrollWeiss1965}.
For further outline of renewal theory and related subjects we
refer to the references
\cite{Shlesinger2017,Gorenflo2007,MontrollWeiss1965,MainardiGorenfloScalas2004,ScherLax1973,KutnerMasoliver1990}.
It is mention worthy that we deal in this paper with causal generalized functions and distributions in the sense of Gelfand and Shilov \cite{GelfangShilov1968}.

We consider a sequence of randomly occurring `events'. Such events can be for instance the jumps of a diffusing particle or failure events in technical systems.
We assume that the events occur at non-negative random times $\it 0 \leq t_1,t_2,\ldots t_n, \ldots , \infty$
where $t=0$ represents the start of the observation.
The random times $t_k$ when events occur are called `arrival times'.
The time intervals between successive events $\it \Delta t_k = t_k-t_{k-1} \geq 0$ are called `waiting times' or `interarrival times'
\cite{MainardiGorenfloScalas2004}.
The random event stream is referred to as a {\it renewal process} if the
waiting time $\it \Delta t$ between successive events is an `independent and identically distributed' (IID) random variable which 
may take any non-negative continuous value. This means
in a renewal process the waiting time $\it \Delta t_k$ between successive events is drawn $\it \forall k$ from the same waiting probability density function
(PDF) $\it \chi(t)$. This distribution function is called
{\it waiting time distribution function} (waiting time PDF) or short {\it waiting time density}\footnote{In the context of random walks where the events
indicate random jumps we also utilize the notion `jump density' \cite{MichelRiascos2019}.}.
The quantity $\it \chi(t){\rm d}t$ indicates the probability that an event occurs at time $\it t$
(within $\it [t,t+{\rm d}t]$).

We can then write
the probability $\it \Psi(t)$ that the waiting time for the first event is $\it \Delta t \leq t$ or equivalently
that {\it at least one event occurs} in the time interval $\it [0,t]$ as
\beq
 \label{waitintimedis}
\Psi(t) = \int_0^t \chi(\tau){\rm d}\tau ,\hspace{0.5cm} t\geq 0 , 
\hspace{0.5cm} \lim_{t\rightarrow\infty} \Psi(t) =1-0
\eeq
with the obvious initial condition $\it \Psi(t=0)=0$. The distribution (\ref{waitintimedis}) in the context of lifetime models often is also called
`failure probability' \cite{MainardiGorenfloScalas2004}.
From this relation follows that
the waiting time density $\it \chi(t)$ is a normalized PDF.
In classical renewal theory the waiting time PDF was assumed to be exponential $\it \chi(t)=\xi e^{-\xi t}$ ($\it \xi >0$) which leads as
we will see later to Markovian memoryless Poisson type processes.

The waiting time PDF has physical dimension $\it sec^{-1}$ and
the cumulative distribution (\ref{waitintimedis}) indicates a dimensionless probability.
Another quantity of interest is the so called 
`survival probability' $\it \Phi^{(0)}(t)$ defined as
\beq
\label{survie}
\Phi^{(0)}(t) = 1-\Psi(t) = \int_t^{\infty} \chi(\tau){\rm d}\tau 
\eeq
which indicates the (dimensionless) probability that no event has occurred within $\it [0,t]$, i.e. in a random walk the probability that
the walker at time $\it t$ still is waiting on its departure site. Of further interest is the PDF 
of the arrival of $\mathit n$ jump events which we denote by $\it \chi^{(n)}(t)$ 
($\it \chi^{(n)}(t){\rm d}t$ being the probability that the $\it n$th jump is performed at time $\it t$).
Since the events are IID we can establish the recursion
\beq
\label{nstepdensityrec}
\chi^{(n)}(t) = \int_0^t \chi^{(n-1)}(\tau)\chi(t-\tau){\rm d}\tau ,\hspace{1cm} \chi^{(0)}(t)=\delta(t) 
\eeq
and with $\it \chi^{(1)}(t) =\chi(t)$.
Thus the PDF for the arrival of the $\it n$th event is given by the $\it n-1$ fold convolution of $\it \chi(t)$ with itself, namely
\beq
 \label{probofnjumps}
 \chi^{(n)}(t) = \int_0^{\infty}\ldots \int_0^{\infty}\chi(\tau_1)\ldots 
 \chi(\tau_n)\delta\left(t-\sum_{j=1}^n\tau_j\right){\rm d}\tau_1\ldots {\rm d}\tau_n ,\hspace{0.5cm} t>0, \hspace{0.5cm} n=1,2,\ldots  .
\eeq
In this relation we have assumed that the waiting time PDF is {\it causal}, i.e. $\it \chi(t)$ is non-zero only for $\it t\geq 0$.
For an outline of causal distributions and some of their properties especially Laplace transforms, see Appendix \ref{causal}.
The probability that $\it n$ events happen within time interval $\it [0,t]$ then 
can be written as 
\beq
 \label{nstepprobabilityuptot}
 \begin{array}{l} 
\ds  \Phi^{(n)}(t) = \int_0^t\left(1-\Psi(t-\tau)\right)\chi^{(n)}(\tau){\rm d}\tau = \int_0^t \Phi^{(0)}(t-\tau)\chi^{(n)}(\tau){\rm d}\tau , \hspace{0.5cm} 
n=0,1,2, \dots \\ \\ 
 \ds \hspace{0.5cm} = \int_0^t \Phi^{(n-1)}(t-\tau)\chi(\tau){\rm d}\tau .
 \end{array}
\eeq
This convolution takes into account that the $\it n$th event may happen at a time $\it \tau < t$ and 
no further event is taking place during $\it t-\tau$ with survival probability $\it \Phi^{(0)}(t)$
where $\it 0\leq \tau \leq t$. The distribution $\it \Phi^{(n)}(t)$ are dimensionless probabilities whereas the PDFs 
$\it \chi^{(n)}(t)$ have 
physical dimension of $\it sec^{-1}$.
It is especially instructive to consider all these convolution relations in the Laplace domain. We then obtain with (\ref{probofnjumps})
the convolution relation
\beq
 \label{the-rule}
 \begin{array}{l}
 \ds {\tilde \chi}^{(n)}(s) = \int_0^{\infty}\ldots \int_0^{\infty}\chi(\tau_1)\ldots 
 \chi(\tau_n)e^{-st} \delta\left(t-\sum_{j=1}^n\tau_j\right){\rm d}\tau_1\ldots {\rm d}\tau_n =\left\{ \int_0^{\infty} \chi(t)e^{-st}{\rm d}t\right\}^n \\ \\
 \ds \hspace{0.5cm} =  ({\tilde \chi}(s))^n ,\hspace{1cm} n=0,1,2,\ldots
 \end{array}
\eeq
where $\it {\tilde \chi}^{(0)}(s) = 1$ indeed recovers $\it \chi^{(0)}(t)=\delta(t)$ for $\it n=0$. 
This relation also shows that
the density of $\it n$ events $\it \chi^{(n)}(t)$ is normalized, namely
\beq
\label{normalizationconchi}
{\tilde \chi}^{(n)}(s)|_{s=0} = 1 
\eeq
as a consequence of the normalization of the waiting time PDF $\it \chi(t)$.
Now in view of (\ref{waitintimedis}) and (\ref{survie}) it is straightforward to obtain the Laplace transforms
\beq 
\label{Laplcezero}
{\tilde \Psi}(s) =\frac{{\tilde \chi}(s)}{s} ,\hspace{1cm}  {\tilde \Phi}^{(0)}(s) =\frac{1}{s}-{\tilde \Psi}(s) = \frac{1-{\tilde \chi}(s)}{s}
\eeq
thus the Laplace transform of the probability distribution (\ref{nstepprobabilityuptot}) for $\it n$ events is given by
\beq
\label{narrivalsLaplace}
{\tilde \Phi}^{(n)}(s) = {\tilde \Phi}^{(0)}(s) ({\tilde \chi}(s))^n = \frac{1-{\tilde \chi}(s)}{s} ({\tilde \chi}(s))^n ,
\hspace{0.5cm} n=0,1,2,\ldots
\eeq
For a brief demonstration of further general properties of renewal processes it 
is convenient to introduce the following generating function
\beq
\label{generfu}
G(t,v)= \sum_{n=0}^{\infty} v^n \Phi^{(n)}(t)
\eeq
and its Laplace transform\footnote{We denote $\it {\tilde f}(s) = {\cal L}\{f(t)\}$ the Laplace transform of $\it f(t)$ and by 
$\it {\cal L}^{-1}\{\ldots\}$ Laplace inversion, 
see Appendix \ref{causal} for further details.}
\beq
\label{generfulaplace}
{\tilde G}(s,v) = {\cal L} \{G(t,v)\}= \sum_{n=0}^{\infty} v^n {\tilde \Phi}^{(n)}(s).
\eeq
Taking into account (\ref{narrivalsLaplace}) together with the obvious property 
$\it |{\tilde \chi}(s)| \leq  |{\tilde \chi}(s=0)| = 1$  we get
for (\ref{generfulaplace}) a geometric series
\beq
\label{geoLplacegenfu}
{\tilde G}(s,v) = \frac{1-{\tilde \chi}(s)}{s} \sum_{n=0}^{\infty} v^n ({\tilde \chi}(s))^n =  \frac{1-{\tilde \chi}(s)}{s} \frac{1}{1-v{\tilde \chi}(s)} 
\eeq
converging for $\it |v ({\tilde \chi}(s))|<1$, i.e. for $\it |v| \leq 1$ if $\it s \neq 0 $ and  $\it |v| < 1$ for $\it s=0$. 
We directly observe in this relation the
normalization condition
\beq
\label{invLaplcace}
{\cal L}^{-1}\{ {\tilde G}(s,v)|_{v=1}\} = {\cal L}^{-1}\left\{ \frac{1}{s}\right\} = G(t,1) = \sum_{n=0}^{\infty} \Phi^{(n)}(t) = 1 ,\hspace{1cm} t>0.
\eeq
The generating function is often useful for the explicit determination of the 
$\it \Phi^{(n)}(t)$, namely
\beq
\label{generdet}
 \Phi^{(n)}(t) = \frac{1}{n!} \frac{d^n}{dv^n} G(t,v)\Big|_{v=0} = {\cal L}^{-1}\left\{ \frac{1-{\tilde \chi}(s)}{s} ({\tilde \chi}(s))^n   \right\}
\eeq
where the Laplace transform of this relation recovers by accounting for (\ref{geoLplacegenfu}) again the expression (\ref{narrivalsLaplace}).

Of further interest is the expected number of events $\it {\bar n}(t)$ that are taking place within the time 
interval $\it [0,t]$. This quantity can be obtained from 
the relation
\beq
\label{meantevents}
{\bar n}(t) = \sum_{n=0}^{\infty}  n \Phi^{(n)}(t) =  \frac{d}{dv} G(t,v)\Big|_{v=1} = {\cal L} ^{-1} \left\{\frac{d}{dv} {\tilde G}(s,v)\Big|_{v=1}\right\} = {\cal L} ^{-1}
\left\{ \frac{{\tilde \chi}(s)}{s (1-{\tilde \chi}(s))} \right\}.
\eeq 
\subsection{POISSON PROCESS}
\label{Poisson}
Before we pass on to non-classical generalizations, it appears instructive to recall 
some properties of the classical variant which is the `Poisson renewal process' (compound Poisson process) \cite{Feller1968}.
In this process the waiting time PDF has exponential form
\beq
\label{Poisson-class}
\chi_P(t) =\xi e^{-\xi t} \Theta(t) ,\hspace{1cm} \xi > 0
\eeq
where $\it \xi$ is a characteristic constant with physical dimension $\it sec^{-1}$ where $\it \xi^{-1}$ defines a characteristic time scale in the process. 
With the Heaviside $\it \Theta(t)$-function we indicate here that (\ref{Poisson-class}) is a causal distribution\footnote{We often skip $\it \Theta(t)$ 
when there is no time derivative involved.}. 
We see that (\ref{Poisson-class}) is a normalized PDF which has the Laplace transform
\beq
 \label{LaplcePoisson}
 {\tilde \chi}_P(s) = \xi \int_0^{\infty}e^{-s t}  e^{-\xi t} {\rm d}t =\frac{\xi}{\xi+s} 
\eeq
where $\it {\tilde \chi}_P(s=0)=1$ reflects normalization of waiting time PDF (\ref{Poisson-class}).
Then we get straightforwardly the failure and survival probabilities, respectively
\beq
\label{failurePoisson}
\Psi_P(t) = 1-e^{-\xi t} ,\hspace{1cm} \Phi^{(0)}_P(t) =1-\Psi_P(t) = e^{-\xi t} .
\eeq
Also the generating function can be written down directly as
\beq
 \label{GenerPoisson}
 {\tilde G}_P(s,v) = \frac{1}{\xi+s}\sum_{n=0}^{\infty} \frac{(\xi v)^n }{(\xi+s)^n} = \frac{1}{\xi(1-v)+s}  \hspace{1cm} \Re \{s\} > \xi ,
\eeq
thus
\beq
\label{genfutimedomain}
G_P(t,v) = e^{-\xi(1-v) t} .
\eeq
By using (\ref{generdet}) we obtain then for the probability of $\it n$ events
\beq
\label{Poissondistribution}
\Phi^{(n)}_P(t) =  \frac{1}{n!} \frac{d^n}{dv^n} e^{(v-1)\xi t}\,\,\Big|_{v=0}  =  \frac{(\xi t)^n}{n!} e^{-\xi t} ,\hspace{0.5cm} n=0,1,2,\dots  ,\hspace{0.5cm} t\geq 0
\eeq 
which is the {\it Poisson distribution}. Therefore the renewal process generated
by an IID exponential waiting time PDF (\ref{Poisson-class}) is referred to as
{\it Poisson renewal process} or also {\it compound Poisson process}. 
This process is the classical proto-example of renewal process \cite{MainardiGorenfloScalas2004,Feller1968} 
(and see the references 
therein).
We further mention in view of Eq. (\ref{meantevents}) that the average number of events $\it {\bar n}(t)$ taking place within $\it [0,t]$
is obtained as
\beq
\label{averagenumberPoisson}
{\bar n}_p(t)= \frac{d}{dv}G_P(t,v)\Big|_{v=1} = \frac{d}{dv} e^{(v-1) \xi t}\Big|_{v=1} = \xi t ,\hspace{1cm} t\geq 0.
\eeq
In a Poisson renewal process the  expected number of arrivals increases linearly in time. The exponential decay in the distributions related to the Poisson process
make this process memoryless with the Markovian property \cite{MichelRiascos2019,MainardiGorenfloScalas2004}.

\section{\small FRACTIONAL POISSON PROCESS}
\label{fractionalPoisson}
In anomalous diffusion one has for the average number of arrivals instead of the linear behavior (\ref{averagenumberPoisson}) a
power law $\it \sim t^{\beta}$ with $\it 0<\beta<1$  \cite{MetzlerKlafter2000,MichelRiascos2019,MichelitschRiascosGFPP2019}, among others. 
To describe such anomalous power-law behavior a
`fractional generalization' of the classical Poisson renewal process was introduced and analyzed by Laskin 
\cite{Laskin2003,Laskin2009} and others
\cite{RepinSaichev2000,MainardiGorenfloScalas2004,BeghinOrsinger2009}. 
The {\it fractional Poisson renewal process} can be defined by a waiting time PDF with the Laplace transform
\beq
\label{fractgen}
 {\tilde \chi}_{\beta}(s) = \frac{\xi}{s^{\beta}+\xi} ,\hspace{1cm} \xi >0, \hspace{1cm} 0 < \beta \leq 1 .
 \eeq
The fractional Poisson process introduces long-time memory effects with non-Markovian features. 
We will come back to these issues later.
The constant $\it \xi$ has here physical dimension $\it sec^{-\beta}$ defining a characteristic time 
scale in the fractional Poisson process. 
For $\it \beta=1$ the fractional Poisson process recovers the standard Poisson process outlined in the previous section.
The waiting time density of the fractional Poisson process is then defined by
\beq
\label{fracwaitibngtimePDF}
\chi_{\beta}(t) = {\cal L}^{-1}\left\{ \frac{\xi}{s^{\beta}+\xi}\right\} = {\cal L}^{-1} \left\{\xi s^{-\beta}\frac{1}{1+\xi s^{-\beta}} \right\}.
\eeq
In order to evaluate the inverse Laplace transform it useful to expand $\it (1+\xi s^{-\beta})^{-1}$ into a geometric series 
with respect to $\it \xi s^{-\beta}$ which converges for
$\it s=\sigma+i\omega$ with $\it \sigma = \Re\{s\} > \xi^{\frac{1}{\beta}} $ for all $\it \omega$. Doing so we obtain
\beq
\label{mittag-leffler-density-lap}
\chi_{\beta}(t) = \sum_{m=0}^{\infty} (-1)^m \xi^{m+1}  {\cal L}^{-1}\{s^{-\beta(m+1)}\} ,\hspace{1cm} 0<\beta \leq 1 ,\hspace{1cm} \Re\{s\} > \xi^{\frac{1}{\beta}} .
\eeq
Taking into account the inverse Laplace transform\footnote{In this relation in the 
sense of generalized functions we can include the value $\it \mu=0$ as as he limit 
$\it \lim_{\mu\rightarrow 0+} \frac{t^{\mu-1}}{\Gamma(\mu)} =\delta(t)$ 
\cite{GelfangShilov1968}.} $\it {\cal L}^{-1}\{ s^{-\mu}\} =\Theta(t) \frac{t^{\mu-1}}{\Gamma(\mu)}$ where $\it \mu>0$
(See also Appendix \ref{causal} for the discussion of some properties). 
We obtain then for (\ref{mittag-leffler-density-lap}) 
\cite{Laskin2003,MichelRiascos2019,MainardiGorenfloScalas2004}
\beq
\label{Mittag-Leffler-density}
\begin{array}{l} 
\ds \it \chi_{\beta}(t) =  \xi t^{\beta-1} \sum_{m=0}^{\infty} \frac{(- \xi t^{\beta})^m}{\Gamma(\beta m +\beta)} ,\hspace{1cm} 0<\beta \leq 1 ,\hspace{1cm} t >0 \\ \\
\ds \it \hspace{0.5cm} =  \xi t^{\beta-1} E_{\beta,\beta}(-\xi t^{\beta}) =  \frac{d}{dt}(1-E_{\beta}(-\xi t^{\beta})) 
\end{array}
\eeq 
where in this relation we introduced the generalized Mittag-Leffler function $\it E_{\beta,\gamma}(z)$ and the standard Mittag-Leffler function $\it E_{\beta}(z)$
defined in the Appendix \ref{causal} by Eqs. (\ref{genmiLeff}) and (\ref{mittag-le}), respectively. The waiting time PDF of the fractional Poisson process also is 
referred to as Mittag-Leffler density and was introduced first by Hilfer and Anton \cite{HilferAnton1995}.
It is now straightforward to obtain in the same way the survival probability for the fractional Poisson process, namely (See also Eq. (\ref{Laplcezero}))
\beq
\label{fracPoissonsurv}
\Phi^{(0)}_{\beta}(t)= {\cal L}^{-1}\left\{\frac{s^{\beta-1}}{s^{\beta}+\xi}\right\} =  E_{\beta}(-\xi t^{\beta}) ,\hspace{1cm}  0<\beta \leq 1 .
\eeq
The generating function (\ref{generfu}) is then by accounting for (\ref{fracPoissonsurv}) obtained as
\beq
\label{genfufrac}
G_{\beta}(t,v) = {\cal L}^{-1}\left\{\frac{s^{\beta-1}}{\xi(1-v)+s^{\beta}}\right\} = E_{\beta}(-\xi(1-v)t^{\beta}) ,\hspace{1cm} t\geq 0.
\eeq
For $\it v=1$ this relation takes $\it G(t,1)=\Theta(t) =1 $ ($\it t\geq 0$) and for $\it \beta=1$ the Poisson exponential 
(\ref{genfutimedomain}) is recovered.
The probability for $\it n$ arrivals within $\it [0,t]$ is then with relation (\ref{generdet}) obtained as 
\beq
\label{narrivalsfracPoisson}
\Phi^{(n)}_{\beta}(t) = \frac{1}{n!} \frac{d^n}{dv^n}  E_{\beta}((v-1)\xi t^{\beta}) \, \Big|_{v=0} = 
\frac{(\xi t^{\beta})^n}{n!} \sum_{m=0}^{\infty}\frac{(m+n)!}{m!} \frac{(-\xi t^{\beta})^m}{\Gamma(\beta(m+n)+1)} ,\hspace{1cm} 0<\beta\leq 1 .
\eeq
This distribution is called the {\it fractional Poisson distribution} and is of utmost importance in fractional dynamics, 
generalizing the Poisson
distribution (\ref{Poissondistribution}) \cite{Laskin2003,Laskin2009}. For $\it \beta=1$ 
the fractional Poisson distribution (\ref{narrivalsfracPoisson}) turns into
the classical Poisson distribution (\ref{Poissondistribution}). 
We directly confirm the normalization of the fractional Poisson distribution by the relation
\beq
\label{normalizationfractional}
\sum_{n=0}^{\infty} \Phi^{(n)}_{\beta}(t) =  \sum_{n=0}^{\infty} \frac{1}{n!} \frac{d^n}{dv^n}  E_{\beta}((v-1)\xi t^{\beta}) \, \Big|_{v=0} = 
E_{\beta}((-1+1)\xi t^{\beta}) = E_{\beta}(0) = 1.
\eeq
We notice that for $\beta=1$ the Mittag-Leffler function becomes the 
exponential $\it E_{1}(-\xi t) = e^{-\xi t}$ thus the
distributions of the standard Poisson process of last section are then reproduced. 
It is worthy to consider the distinct behavior of the fractional Poisson process for large observation times. 
To this end let us expand Laplace transform (\ref{fractgen})
for $\it |s|$ small which governs the asymptotic behavior for large times
\beq
 \label{frac-exp}
 {\tilde \chi}_{\beta}(s) = \left(1+\frac{s^{\beta}}{\xi}\right)^{-1} = \sum_{m=0}^{\infty}
 (-1)^m \xi^{(-m)}s^{m\beta} = 1-\frac{1}{\xi} s^{\beta}+\ldots 
 \eeq
which yields as asymptotically for large observation times for $\it 0<\beta<1$, $\alpha>0$ 
fat-tailed behavior\footnote{Note that
$\it -\Gamma(-\beta)=\beta^{-1}\Gamma(1-\beta) >0$.}
 \beq
  \label{tlarge-fat-tailed-fracPoisson}
  \chi_{\beta}(t) \approx -\frac{1}{\xi \Gamma(-\beta)}t^{-\beta-1} ,
  \hspace{0.5cm} 0<\beta< 1 ,\hspace{0.5cm} \alpha>0 ,\hspace{0.5cm}
  t\rightarrow\infty .
 \eeq
The fat-tailed behavior $\it \chi_{\beta}(t) \sim t^{-\beta-1}$ is a characteristic power-law feature of the 
fractional Poisson renewal process reflecting the non-locality in time that produces Laplace 
transform (\ref{frac-exp}) within the fractional index range $\it 0<\beta<1$. As a consequence of the fat-tailed behavior
for $\it 0<\beta<1$ extremely long waiting times occur thus the fractional Poisson process is non-Markovian exhibiting long-time memory effects \cite{MichelRiascos2019,MainardiGorenfloScalas2004}.  
\\

Further of interest is the power-law tail in the fractional Poisson distribution. We obtain this behavior
by considering the lowest power in their Laplace transform, namely
\beq
\label{long-range_power}
\begin{array}{l} 
\it \ds \Phi_{\beta}^{(n)}(t) = 
{\cal L}^{-1} 
\left\{\frac{1}{s}\left( (1+\xi^{-1}s^{\beta})^{-n} - (1+\xi^{-1}s^{\beta})^{-n-1} \right)\right\}
\approx {\cal L}^{-1}\left\{ \frac{s^{\beta-1} }{\xi} \right\} \\ \\ \hspace{0.5cm} \it \ds \approx \frac{(t\xi^{\frac{1}{\beta}})^{-\beta}}{ \Gamma(1-\beta)}
\hspace{0.5cm} 0<\beta<1 ,\hspace{0.5cm} n=0,1,\dots ,\hspace{1cm} t\xi^{\frac{1}{\beta}} \rightarrow \infty.
\end{array}
\eeq
The fractional Poisson distribution exhibits for large (dimensionless) observation times $\it t\xi^{\frac{1}{\beta}}\rightarrow \infty$ universal
$\it t^{-\beta}$ power-law behavior independent of the arrival number $\it n$. We will come back subsequently to this important issue.

\section{\small GENERALIZATION OF THE FRACTIONAL POISSON PROCESS}
\label{GFPP}
In this section our aim is to develop a renewal process which is a 
generalization of the fractional Poisson process of previous section.
The waiting time PDF of this
process has
the Laplace transform 
\beq
 \label{jump-gen-fractional-laplacetr}
 {\tilde \chi}_{\beta,\alpha}(s) = \frac{\xi^{\alpha}}{(s^{\beta}+\xi)^{\alpha}} ,\hspace{0.5cm} 
 0<\beta\leq 1 ,\hspace{0.5cm} \alpha >0 , \hspace{1cm} \xi >0 .
\eeq
This process was first introduced by Cahoy and Polito \cite{PolitoCahoy2013}.
We referred the renewal process defined by (\ref{jump-gen-fractional-laplacetr}) to as 
the {\it generalized fractional Poisson process} (GFPP) \cite{MichelRiascos2019,MichelitschRiascosGFPP2019}. 
The characteristic dimensional constant $\it \xi$ in
(\ref{jump-gen-fractional-laplacetr}) has as in the fractional Poisson process physical dimension 
$\it \sec^{-\beta}$ and defines 
a characteristic time scale. The GFPP contains further two index parameters
$\it \alpha >0$ and $\it 0<\beta \leq 1$. The advantage of generalizations such as the GFPP is that they offer a larger
parameter space allowing to greater flexibility in adapting to real-world situations.
The GFPP recovers for $\it \alpha=1$, $\it 0<\beta<1$ the above described fractional Poisson process and for
$\it \alpha=1$, $\it \beta=1$ the standard Poisson process and for $\it \beta=1$, $\it \alpha>0$ the so called 
(generalized) Erlang process
where $\it \alpha$ is allowed to take positive integer or non-integer values \cite{MichelRiascos2019}.
The waiting time density of the GFPP is then obtained as (See also Ref. \cite{PolitoCahoy2013})
\beq
 \label{Laplainv}
 \begin{array}{l}
 \it
\ds  \chi_{\beta,\alpha}(t)  = \xi^{\alpha} \sum_{m=0}^{\infty} (-1)^m \frac{(\alpha)_m}{m!} 
 \xi^m {\cal L}^{-1}\{s^{-\beta (m+\alpha)}\} ,\hspace{0.25cm} t>0, \hspace{0.2cm} \sigma=\Re\{s\}>\xi^{\frac{1}{\beta}}, \hspace{0.2cm} 0<\beta\leq 1, 
 \hspace{0.2cm} \alpha >0\\ \\ \it
 \ds \hspace{0.5cm}  = \xi^{\alpha} t^{\beta\alpha-1} \sum_{m=0}^{\infty}
 \frac{(\alpha)_m}{m!}\frac{(-\xi t^{\beta})^m}{\Gamma(\beta m+ \alpha\beta)} = 
 \xi^{\alpha} t^{\beta\alpha-1} E_{\beta,\alpha\beta}^{\alpha}(-\xi t^{\beta}).
 \end{array} 
 \eeq
In this expression is introduced a generalization of the Mittag-Leffler 
function which was first described by Prabhakar \cite{Prabhakar1971} and is defined by
\beq
 \label{genmittag-Leff}
 E_{a,b}^c(z) = \sum_{m=0}^{\infty}
 \frac{(c)_m}{m!}\frac{z^m}{\Gamma(am + b)} ,\hspace{0.5cm} \Re\{a\} >0 ,
 \hspace{0.5cm} \Re\{b\} > 0 ,\hspace{0.5cm} c,\, z \in \mathbb{C}.
\eeq
In the Prabhakar-Mittag-Leffler function (\ref{genmittag-Leff}) and in the expansion (\ref{Laplainv}) we introduced 
the Pochhammer symbol $\it (c)_m$ which is defined as \cite{Mathai2010}
\beq
\label{Pochhammer}
(c)_m =  
\frac{\Gamma(c+m)}{\Gamma(c)} = \left\{\begin{array}{l} 1 ,\hspace{1cm} m=0 \\ \\ 
                                                            
     c(c+1)\ldots (c+m-1) ,\hspace{1cm} m=1,2,\ldots \end{array}\right .
\eeq
Despite $\it \Gamma(c)$ is singular at $\it c=0$ the Pochhammer symbol can be defined also
for $\it c=0$ by the limit $\it (0)_m= \lim_{c\rightarrow 0+}(c)_m=\delta_{m0}$ which is also 
fulfilled by the right-hand side of (\ref{Pochhammer}). Then $\it (c)_m$ is defined
for all $\it c \in \mathbb{C}$ thus we have $\it E_{a,b}^0(z) =1$. 
The series (\ref{genmittag-Leff}) converges absolutely in the entire complex $\it z$-plane. 
\begin{figure*}[!t]
\begin{center}
\includegraphics*[width=1.0\textwidth]{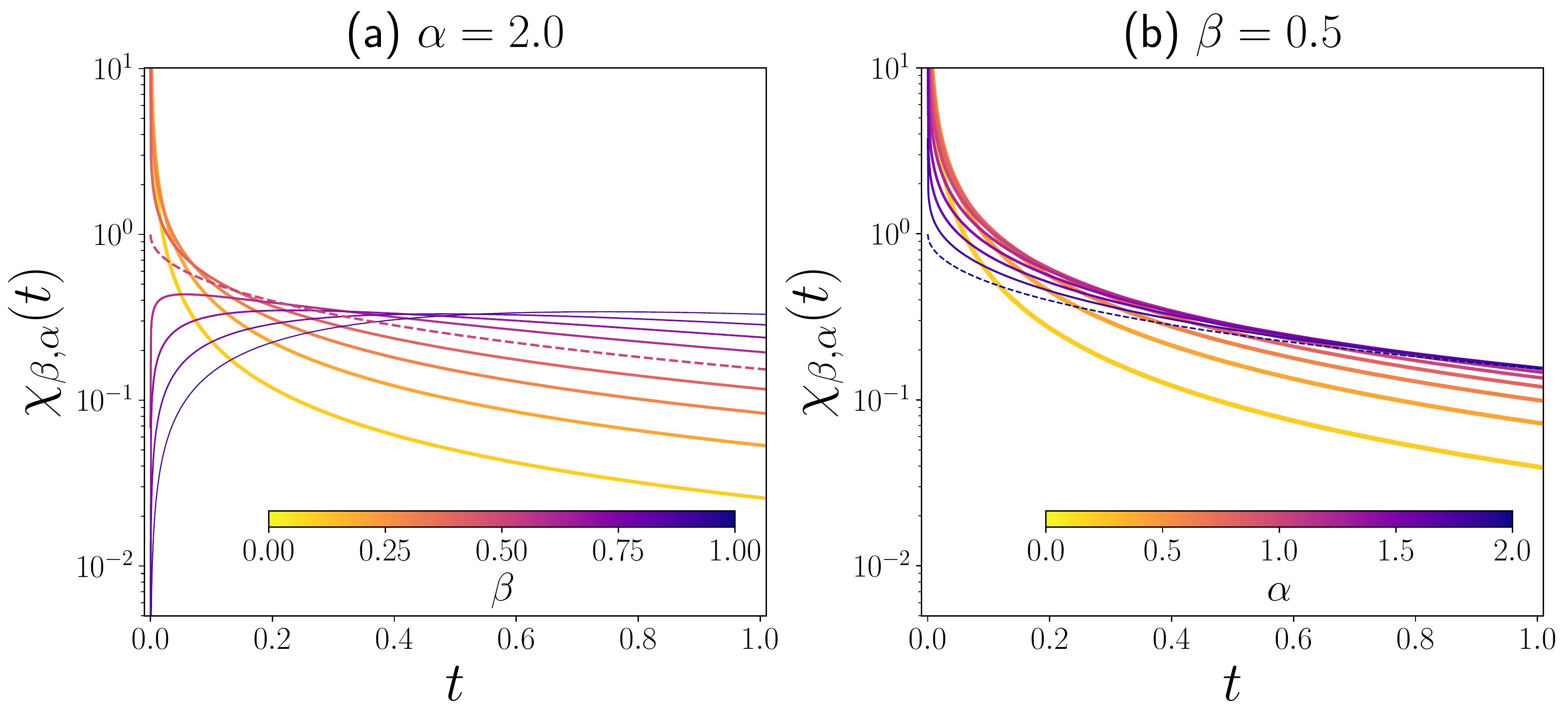}
\end{center}
\vspace{-5mm}
\caption{\label{Fig_1} The waiting time density  density $\chi_{\beta,\alpha}(t)$ as a function of $t$. We explore the results for (a) $\alpha=2.0$  for different values  $0< \beta \leq 1$ and (b) $\beta=0.5$  for   $0< \alpha \leq 2$ (in each case, the parameters are codified in the colorbar). Results were obtained numerically using $\xi=1$ and Eqs. (\ref{Laplainv}) and (\ref{genmittag-Leff}). We depict with dashed lines the case when $\alpha\beta=1$.}
\end{figure*} 
The Prabhakar-Mittag-Leffler function (\ref{genmittag-Leff}) and related problems were
analyzed by several authors 
\cite{Mathai2010,ShulkaPrajabati2007,HauboldMathaiSaxena2011,GaraGarrappa2018,GarraGorenfloPolito2014}.

In Figure \ref{Fig_1}(a) is drawn the waiting time PDF of Eq. (\ref{Laplainv}) for a fixed value of $\it \alpha=2$ and 
variable $\beta$ in the admissible range $\it 0<\beta \leq 1$. The waiting time PDF exhibits for $\it t $ small the power-law behavior
$\it \chi_{\beta,\alpha}(t) \approx \frac{\xi^{\alpha}}{\Gamma(\alpha\beta)} t^{\alpha\beta-1}$ (corresponding to the zero order in the expansion 
(\ref{Laplainv})) with two distinct regimes: For $\it \alpha\beta < 1$ the waiting time PDF becomes singular at $\it t=0$ corresponding to
`immediate' arrivals of the first event. For $\it \alpha\beta=1$ the jump density takes the constant value $\it \chi_{\alpha^{-1},\alpha}(t=0) = \xi^{\alpha}$
whereas for $\it \alpha\beta>0$ the waiting time density $\it \chi_{\beta,\alpha}(t=0) =0$ tends to zero as $\it t \rightarrow 0$ 
where the waiting times become longer the larger $\alpha\beta$.

In Figure \ref{Fig_1}(b) we depict the behavior of the waiting time PDF for fixed $\it \beta=0.5$ and $\it 0<\alpha\leq 2$ thus $\it \alpha\beta \leq 1$. It can be seen that
the smaller $\it \alpha\beta$ the more narrowly the waiting time PDF is concentrated at small $\it t$-values close to $\it t=0$. 
This behavior can also be identified in view of Laplace transform
(\ref{jump-gen-fractional-laplacetr}) which takes in the limit 
$\it \alpha\rightarrow 0$ the value 
$\it \lim_{\alpha\rightarrow 0} {\tilde \chi}_{\beta,\alpha}(s) =1$ thus $\it \chi_{\beta,0+}(t) ={\cal L}^{-1}(1) = 
\lim_{\alpha\rightarrow 0} \frac{\xi^{\alpha}}{\Gamma(\alpha\beta)} t^{\alpha\beta-1} =\delta(t)$ exhibits the shape of a Dirac $\it \delta$-distribution peak.

Now our goal is to determine the generalization of the fractional Poisson distribution (\ref{narrivalsfracPoisson})
which is determined by Eq. (\ref{narrivalsLaplace}) with (\ref{jump-gen-fractional-laplacetr}), namely
\beq
\label{generalizeedFractPoissonDis}
\Phi^{(n)}_{\beta,\alpha}(t) ={\cal L}^{-1}\left\{ 
\frac{1}{s}\left({\tilde \chi}^n_{\beta,\alpha}(s)-{\tilde \chi}^{n+1}_{\beta,\alpha}(s)\right)\right\}= 
{\cal L}^{-1}\left\{ 
\frac{1}{s}\left({\tilde \chi}_{\beta,n\alpha}(s)-{\tilde \chi}_{\beta,(n+1)\alpha}(s)\right) \right\}
,\hspace{5mm} n=0,1,2,\ldots 
\eeq
where it is convenient to utilize 
$\it {\tilde \chi}^n_{\beta,\alpha}(s) =  {\tilde \chi}_{\beta,n\alpha}(s) =
\frac{\xi^{n\alpha}}{(\xi+s^{\beta})^{n\alpha}}$, i.e. to replace $\alpha \rightarrow n\alpha$ in the expression 
(\ref{Laplainv}).
We then obtain for the probability for $\it n$ arrivals within $\it [0,t]$ the expression
\beq
 \label{Generalized-Fractional-Poisson-Distribution}
   \ds \Phi^{(n)}_{\beta,\alpha}(t)  = 
   \xi^{n\alpha} t^{n\alpha\beta} \left( E^{n\alpha}_{\beta,(n\alpha\beta+1)}(-\xi t^{\beta}) 
 - \xi^{\alpha} t^{\alpha\beta} E^{(n+1)\alpha}_{\beta,((n+1)\alpha\beta+1)}(-\xi t^{\beta}) \right).
\eeq
We refer this distribution to as the `{\it generalized fractional Poisson distribution} (GFPD)' 
\cite{MichelRiascos2019,MichelitschRiascosGFPP2019}. This distribution was also obtained by Cahoy and Polito \cite{PolitoCahoy2013}. 
For $\it \alpha=1$ and $\it 0<\beta<1$ the 
GFPP (\ref{Generalized-Fractional-Poisson-Distribution})
recovers the fractional Poisson distribution (\ref{narrivalsfracPoisson}) and for $\it \alpha=1$, $\it \beta=1$ 
the standard Poisson distribution (\ref{Poissondistribution}), and finally for $\it \alpha>0$ and $\it \beta=1$ 
the Erlang distribution \cite{MichelRiascos2019,MichelitschRiascosGFPP2019}.
\begin{figure*}[!t]
\begin{center}
\includegraphics*[width=1.0\textwidth]{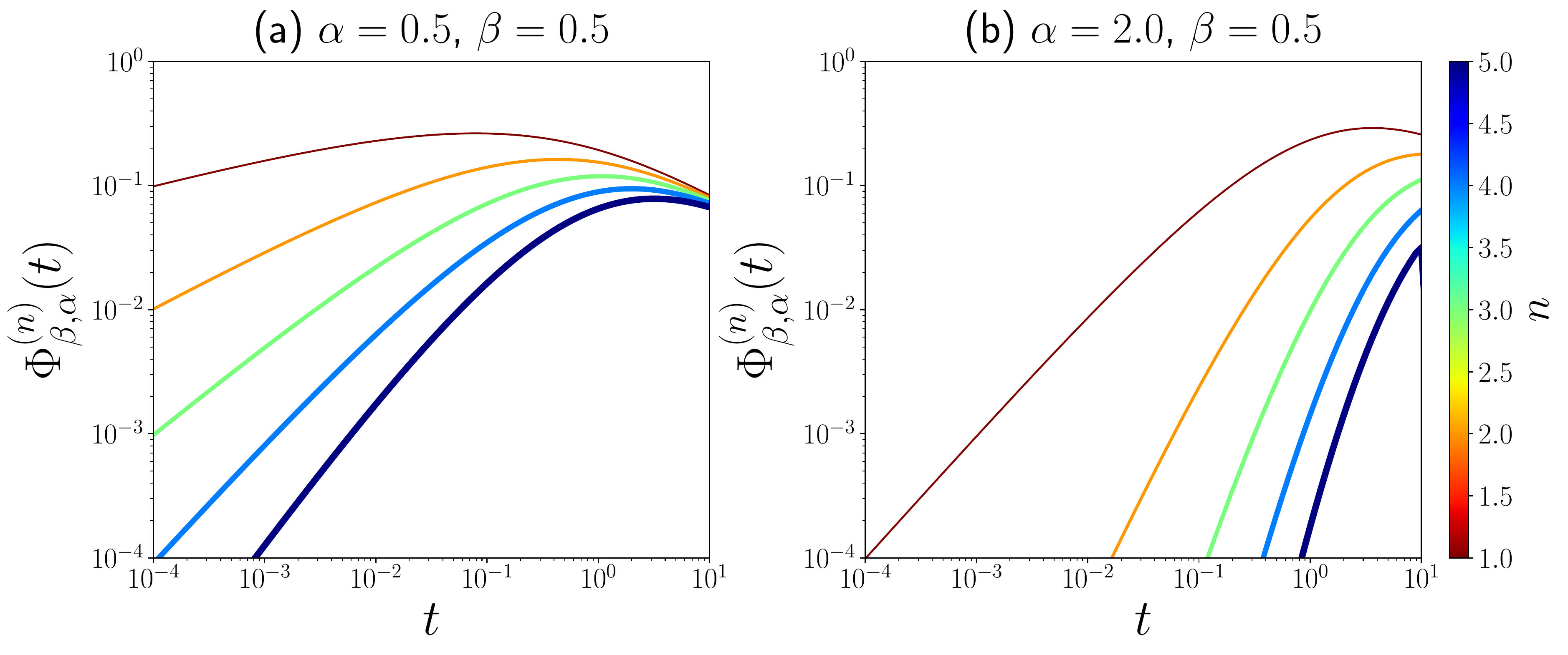}
\end{center}
\vspace{-5mm}
\caption{\label{Fig_2} Probability $\Phi_{\beta,\alpha}^{(n)}(t)$  as a function of $t$ for different $n$. (a) $\alpha=0.5$ and $\beta=0.5$, (a) $\alpha=2.0$ and $\beta=0.5$. In the colorbar we represent $n=1,2,\ldots,5$. The values were obtained numerically using $\xi=1$ with Eq. (\ref{generalizeedFractPoissonDis}). The results for $t\ll 1$ show the power-law relation
$\Phi_{\beta,\alpha}^{(n)}(t)\propto t^{n\alpha\beta}$ in Eq. (\ref{limitsmalltilmes}).
}
\end{figure*} 
For applications in the dynamics in complex systems the asymptotic properties of the GFPD are of interest.
For small (dimensionless) times the GFPD behaves as
\beq
 \label{limitsmalltilmes}
 \Phi^{(n)}_{\beta,\alpha}(t) \approx \frac{(\xi t^{\beta})^{n\alpha}} {\Gamma(n\alpha\beta +1)} ,\hspace{1cm} 
 t\xi^{\frac{1}{\beta}} \rightarrow 0 ,\hspace{1cm} n=0,1,2,\ldots
\eeq
representing the lowest non-vanishing order in (\ref{Generalized-Fractional-Poisson-Distribution}). 
It follows that the GFPD  fulfills the initial condition
\beq
 \label{initialcon}
 \Phi^{(n)}_{\beta,\alpha}(t)\Big|_{t=0} = \delta_{n0} ,
\eeq
reflecting that per construction at $\it t=0$ no event has arrived.
Further of interest is the asymptotic behavior for large (dimensionless) times $\it t\xi^{\frac{1}{\beta}}$.
To this end, let us expand the Laplace transform for small $\it s\rightarrow 0$ 
in (\ref{generalizeedFractPoissonDis}) up to the 
lowest non-vanishing order in $\it \frac{s^{\beta}}{\xi}$ to arrive at
\beq
\label{largetimes}
\Phi^{(n)}_{\beta,\alpha}(t) \approx \frac{\alpha}{\xi} {\cal L}^{-1}\{s^{\beta-1}\}= \frac{\alpha}{\Gamma(1-\beta)}
\left(t \xi^{\frac{1}{\beta}}\right)^{-\beta} ,\hspace{0.5cm} t\xi^{\frac{1}{\beta}} \rightarrow \infty 
, \hspace{0.5cm} n=0 , 1, \dots ,\infty
\eeq
where this inverse power law holds universally for all $\it \alpha>0$ for $\it 0< \beta < 1$ 
and is independent of the number of arrivals $\it n$ recovering the fractional Poisson distribution for $\it \alpha=1$ of Eq. (\ref{long-range_power}). 
We notice that for large (dimensionless) observation times an universal $\it (t\xi^{\frac{1}{\beta}})^{-\beta}$ 
power-law decay occurs which is {\it independent} of the arrival number $\it n$ where
$\it \alpha$ occurs only as a scaling parameter in relation (\ref{largetimes}).
We interpret this behavior as quasi-ergodicity property, i.e. quasi-equal distribution of all `states' $\it n$ for $t\xi^{\frac{1}{\beta}}$ 
large \cite{MichelRiascos2019}.
The fractional exponent $\it -\beta$ further is 
independent of $\it \alpha$ thus the power-law is of the same type as in the fractional Poisson process. 

In the Figure \ref{Fig_2}(a) we have plotted the probabilities $\it \Phi^{(n)}_{\beta,\alpha}(t)$ of Eq. (\ref{Generalized-Fractional-Poisson-Distribution})
for fixed $\it \alpha$ and $\it \beta$ for different arrival numbers $\it n$. One can see that for large times the $\it \Phi^{(n)}_{\beta,\alpha}(t)$ converge to 
the same universal behavior independent of $\it n$ which reflects the asymptotic power-law relation (\ref{largetimes}). 
On the other hand the asymptotic power-law behavior for small $t$ is shown
in Figure \ref{Fig_2}(b) (See also relation (\ref{limitsmalltilmes})). The decay to zero 
$\it \lim_{t\rightarrow 0} \Phi^{(n)}_{\beta,\alpha}(t)\Big|_{t=0} \sim (\xi t^{\beta})^{n\alpha} \rightarrow 0$ ($\it n>0$) becomes the 
more pronounced the larger $\it n$. This behavior also can be interpreted that the higher
$\it n$, the less likely are $\it n$ arrivals to happen within a small time interval of observation.

The GFPP and the fractional Poisson process for large observation times exhibit the same power-law asymptotic feature 
(See again asymptotic relation (\ref{largetimes})). This behavior 
reflects the `asymptotic universality' of the fractional Poisson dynamics, 
the latter was demonstrated in Ref. \cite{GorenfloMainardi2006}. The inverse power-law decay occurring for $\it 0<\beta<1$ 
with fat-tailed waiting time PDF indeed is the source of
non-Markovian behavior with long-time memory. 
In the entire admissible range $\it 0 <\beta \leq 1$ the waiting time PDF $\it \chi_{\beta,\alpha}(t)$ and
survival probability $\it \Phi^{(0)}_{\beta,\alpha}(t)$ maintain their good property of being {\it completely monotonic functions}, i.e.
they fulfill\footnote{See Ref. \cite{MainardiGorenfloScalas2004} 
for a discussion of this issue for the fractional Poisson process.}
\beq
\label{completelymono}
(-1)^n \frac{d^n}{dt^n}f(t) \geq 0 ,\hspace{1cm} n=0,1,2, \dots ,\hspace{1cm} t>0.
\eeq
An analysis of various aspects of completely monotonic functions is performed in our recent works \cite{TMM-APR-ISTE2019,RiascosMichelitsch-et-al-2018},
and see the references therein.

\section{\small CONTINUOUS-TIME RANDOM WALK ON NETWORKS}
\label{CTRWgen}

Having recalled above basic properties of renewal theory\footnote{For further details on renewal theory, see e.g. \cite{Cox1967}.} our goal 
is now to analyze stochastic motions on undirected 
networks and lattices that are governed by the GFPP renewal process. To develop our model we employ the 
continuous-time random walk (CTRW) approach by
Montroll and Weiss \cite{MontrollWeiss1965} (and see also the references \cite{Shlesinger2017,ScherLax1973,KutnerMasoliver1990}).
In the present section our aim is to develop a CTRW model for undirected networks in order to
apply the theory to infinite $\it d$-dimensional integer 
lattices $\it \mathbb{Z}^d$. 

We consider an undirected connected network with $\it N$ nodes which we 
denote with $\it p=1,\ldots ,N$. 
The topology of the network is described by the positive-semidefinite $\it N \times N$
Laplacian matrix which is defined by \cite{TMM-APR-ISTE2019,NohRieger2004,RiascosMateos2014,RiascosMateos2015,TMM-APR-JPhys-A2017,TM-AR-In-memoriamMaugin2018,Michelfrac2014} 
\beq
 \label{Laplacianmat}
 L_{pq}=K_p \delta_{pq}-A_{pq}
\eeq
where $\it {\mathbf A} = (A_{pq})$ denotes the adjacency matrix having elements $\it A_{pq}=1$ 
if a pair $\it pq$ of nodes is connected and $\it A_{pq}=0$ if a pair $\it pq$ is disconnected. 
Further we forbid that nodes are connected with themselves
thus $\it A_{pp}=0$.
In an undirected network adjacency and Laplacian matrices are symmetric. The diagonal 
elements $\it L_{pp}=K_p$ of the Laplacian matrix are referred to as the degrees 
of the nodes $\it p$ counting the number of neighbor nodes of a node $\it p$ with $\it K_p=\sum_{q=1}^NA_{pq}$. 
In order to relate the network topology with random walk features we introduce
the one-step transition matrix $\it {\mathbf W} =(W_{pq})$ which is defined by \cite{TMM-APR-ISTE2019,NohRieger2004}
\beq
 \label{one-step}
  W_{pq} =\frac{1}{K_p}A_{pq}= \delta_{pq}-\frac{1}{K_p}L_{pq} .
\eeq
Generally, the transition matrix is non-symmetric for networks with variable degrees
$\it K_i\neq K_j$ $ (i\neq j)$.
The one-step transition matrix $\it W_{pq}$ defines the conditional probability that 
a random walker which is on node $\it p$ jumps in one step to node $\it q$ where in one step only neighbor nodes
with equal probability $\it \frac{1}{K_p}$ can be reached. We see in definition (\ref{one-step}) 
that the one-step transition matrix $\it \sum_{q=1}^N W_{pq} =1$ and $\it 0 \leq  W_{pq} \leq 1$ and also the $\it n$-step transition matrices $\it {\mathbf W}^n$ 
are (row-)stochastic \cite{TMM-APR-ISTE2019}.

Now let us assume that
each step of the walker from one to another node is associated with a jump event or arrival in a CTRW
with identical transition probability $\it (W_{pq})$ for a step from node $\it p$ to node $\it q$.
We assume the random walker performs IID random steps at
random times $\it 0 \leq t_1,t_2,\ldots t_n,\ldots ,\infty $ in a renewal process 
with IID waiting times $\it \Delta t_k$ where the observation starts at $\it t=0$. 
To this end let us recall some basic relations holding generally, and  then we specify the renewal process to be a GFPP.

Introducing the transition matrix $\it {\mathbf P}(t)= (P_{ij}(t))$ 
indicating the probability to find the walker at time $t$ on node $j$
under the condition that the walker at $\it t=0$ initially was sitting at node $\it i$,
we can write \cite{Cox1967,Gorenflo2010}
\beq
 \label{the-CTRW}
 {\mathbf P}(t) =  {\mathbf P}(0) \sum_{n=0}^{\infty} \Phi^{(n)}(t) {\mathbf W}^n 
\eeq
where we assume here a general initial condition
$\it {\mathbf P}(t)|_{t=0} = {\mathbf P}(0)$ which is fulfilled by accounting for the initial conditions $\it \Phi^{(n)}(t)_{t=0} = \delta_{n0}$.
In this series the $\it \Phi^{(n)}(t)$ indicate the probabilities of $n$ (jump-) events in the renewal process, 
i.e. the probability that the walker performs $\it n$ steps
within $\it [0,t]$ (See Eq. (\ref{nstepprobabilityuptot})), and $\it ({\mathbf W}^n)_{ij}$ indicates the probability that the 
walker in $\it n$ jumps moves from the initial node $\it i$ to node $\it j$.
We observe in view of relation (\ref{invLaplcace}) together with $\it \sum_{j=1}^N({\mathbf W}^n)_{ij} =1$ 
that the normalization condition $\it \sum_{j=1}^NP_{ij}(t)=\sum_{n=0}^{\infty} \Phi^{(n)}(t) = 1$ is fulfilled.
The convergence of series (\ref{the-CTRW}) can be easily proved by using that $\it {\mathbf W}$ has uniquely eigenvalues 
$\it |\lambda_m|\leq 1$  and with $\it |{\tilde \chi}(s)| \leq 1$ \cite{TMM-APR-ISTE2019,MichelRiascos2019}. 
Let us assume that at $\it t=0$ the walker 
is sitting on departure node $\it i$ thus the initial condition is given by $\it P_{ij}(0)=(\delta_{ij})$, then the Laplace transform of (\ref{the-CTRW}) 
writes \cite{MichelRiascos2019}
\beq
 \label{occupationproblaplace}
 {\tilde {\mathbf P}}(s) = \frac{\left(1-{\tilde \chi}(s)\right)}{s} \left\{{\mathbf 1}-
 {\tilde \chi}(s){\mathbf W}\right\}^{-1} 
\eeq
where $\it {\tilde {\mathbf P}}(s)$ has the eigenvalues \cite{MichelRiascos2019}
\beq
 \label{Montroll-Weiss}
{\tilde P}(m,s) = \frac{\left(1-{\tilde \chi}(s)\right)}{s}\frac{1}{(1- \lambda_m{\tilde \chi}(s))} ,\hspace{1cm} m=1,\dots ,N .
\eeq
The $\it \lambda_m$ indicate the eigenvalues of the one-step transition matrix $\it {\mathbf W}$.
This expression is the celebrated {\it Montroll-Weiss formula} \cite{MontrollWeiss1965} and 
occurs in various contexts of physics. 
\section{\small GENERALIZED SPACE-TIME FRACTIONAL DIFFUSION IN $\it \mathbb{Z}^d$}
\label{CTRWGFPP}
In this section our aim is to develop a CTRW which is a random walk subordinated to a GFPP. 
For the random walk on the network we allow long-range jumps which can be described when we replace the Laplacian matrix by its fractional power 
in the one-step transition matrix (\ref{one-step}).
In this way, the walker cannot only jump to connected neighbor nodes, but also to far 
distant nodes in the network 
\cite{MetzlerKlafter2000,MetzlerKlafter2004,TMM-APR-ISTE2019,RiascosMichelitsch-et-al-2018,
RiascosMateos2014,RiascosMateos2015,TMM-APR-JPhys-A2017,TM-AR-In-memoriamMaugin2018,RiascosMateos2012,MichelColletRiascos-etal2017}. 
The model to be developed in this section involves 
both space- and time-fractional calculus.
As an example we consider the infinite $\it d$-dimensional integer lattice $\it \mathbb{Z}^d$.
The lattice points $\it \mathbf{p}=(p_1,\ldots,p_d)$ ($p_j \in \mathbb{Z}_0$) 
represent the nodes where we assume each node is connected to any of its $\it 2d$ neighbor nodes.
The $\it \mathbb{Z}^d$ is an infinite cubic primitive $\it d$-dimensional lattice with lattice-constant one.
In this network any node has identical degree $\it 2d$. 
The one-step transition 
matrix with the elements $\it W^{(\mu)}(\mathbf{p}-\mathbf{q})$ has then the canonic representation \cite{TMM-APR-ISTE2019,TMM-APR-JPhys-A2017}
\beq
\label{fractone-step}
W^{(\mu)}(\mathbf{p}-\mathbf{q}) = 
\frac{1}{(2\pi)^d} \int_{-\pi}^{\pi} {\rm d}k_1 \dots \int_{-\pi}^{\pi}{\rm d}k_d 
e^{i{\mathbf k}\cdot (\mathbf{p}-\mathbf{q})} \lambda^{(\mu)}(\mathbf{k})
\eeq
with the eigenvalues

\beq
 \label{eigvals}
 \lambda^{(\mu)}(\mathbf{k})= 1-\frac{1}{{\cal K}^{(\mu)}}\eta^{\frac{\mu}{2}}(\mathbf{k}) 
 ,\hspace{0.5cm} \eta({\mathbf k}) = 2d-2\sum_{j=1}^d\cos(k_j) , \hspace{1cm} 0 < \mu \leq 2
\eeq
where $\it {\mathbf k} = (k_1,\ldots,k_d)$ denotes the wave-vector with $\it -\pi \leq k_j \leq \pi$. 
One can show that the fractional index $\mu$ is restricted to the interval $\it 0< \mu \leq 2$ 
as a requirement for stochasticity of the one-step transition matrix
\cite{TMM-APR-ISTE2019,RiascosMichelitsch-et-al-2018,TMM-APR-JPhys-A2017}.
In (\ref{eigvals}) the constant $\it {\cal K}^{(\mu)}$ can be conceived as a 
fractional generalization of the degree and is given by the trace of the 
fractional power of Laplacian matrix, namely \cite{TMM-APR-ISTE2019,RiascosMichelitsch-et-al-2018}
\beq
\label{fracdeeg}
\it {\cal K}^{(\mu)}= \frac{1}{N} tr(\mathbf{L}^{\frac{\mu}{2}}) = \frac{1}{N} \sum_{m=1}^N (\eta_m)^{\frac{\mu}{2}} 
\eeq 
where $\it \eta_m$ denote the eigenvalues of the Laplacian matrix (\ref{Laplacianmat}) and in an infinite
network the sum in (\ref{fracdeeg}) has to be performed in the limit $\it N\rightarrow\infty$.
In the $\it \mathbb{Z}^d$ the fractional degree with Eq. (\ref{fracdeeg}) 
is then determined from \cite{TMM-APR-ISTE2019,RiascosMichelitsch-et-al-2018}
\beq
\label{fracdeczd}
{\cal K}^{(\mu)} =\frac{1}{(2\pi)^d} \int_{-\pi}^{\pi}{\rm d}k_1 \dots \int_{-\pi}^{\pi}{\rm d}k_d (\eta({\mathbf k}))^{\frac{\mu}{2}}
\eeq
with the eigenvalues given in Eq. (\ref{eigvals}). It is necessary to account for the fractional degree 
since it plays the role of a normalization factor in the one-step transition matrix (See Eq. (\ref{one-step})).

For the present analysis it is sufficient to consider $\it {\cal K}^{(\mu)}$ 
as a (positive) constant where for $\it \mu=2$ recovers $\it {\cal K}^{(\mu=2)} =2d$ the degree of any node.
The transition matrix (\ref{the-CTRW}) is then 
determined by its Laplace transform (\ref{occupationproblaplace})
which writes in the $\it \mathbb{Z}^d$ as
\beq
 \label{greensfunctionfrac}
  {\tilde P}^{(\mu)}(\mathbf{p-q},s) =
 \frac{\left(1-{\tilde \chi}(s)\right)}{s} \frac{1}{(2\pi)^d} 
 \int_{-\pi}^{\pi}{\rm d}k_1\ldots \int_{-\pi}^{\pi}{\rm d}k_d {\tilde P}({\mathbf k},0)
 \frac{e^{i\mathbf{k}\cdot(\mathbf{p}-\mathbf{q})}}{(1- \lambda^{(\mu)}(\mathbf{k}) {\tilde \chi}(s))} 
\eeq
where $\it {\tilde P}({\mathbf k},0)$ indicates the Fourier 
transform of the initial condition which has the Fourier representation
\beq
\label{initialcondition}
P^{(\mu)}(\mathbf{p}-\mathbf{q},t=0) = P_0(\mathbf{p}-\mathbf{q}) 
= 
\frac{1}{(2\pi)^d} \int_{-\pi}^{\pi} {\rm d}k_1 \dots \int_{-\pi}^{\pi}{\rm d}k_d 
e^{i{\mathbf k}\cdot (\mathbf{p}-\mathbf{q})} {\tilde P}({\mathbf k},0).
\eeq
In order to analyze the diffusive limit of (\ref{greensfunctionfrac}), i.e. its long-wave approximation it will be sufficient to account 
for the eigenvalues (\ref{eigvals}) for $\it k \rightarrow 0$ (where $\it k=|\mathbf{k}|$). 
Then we have with $\it \eta^{\frac{\mu}{2}}(\mathbf{k}) \approx k^{\mu}$
the behavior
\beq
 \label{transitionlaceigs}
 \lambda^{(\mu)}(\mathbf{k}) \approx 1-\frac{1}{{\cal K}^{(\mu)}}k^{\mu} ,\hspace{1cm} 0 < \mu \leq 2 , \hspace{0.5cm}  k\rightarrow 0.
\eeq%
These equations hold so far for space-fractional walks for an arbitrary renewal 
process with waiting time PDF $\it \chi(t) ={\cal L}^{-1}\{ {\tilde \chi}(s)\}$.

Now let us consider a space-fractional walk subordinated to a GFPP with $\it {\tilde \chi}(s)= {\tilde \chi}_{\beta,\alpha}(s)$ of Eq. 
(\ref{jump-gen-fractional-laplacetr}). We denote the corresponding transition matrix of this stochastic motion as 
 $\it {\mathbf P}^{(\mu)}_{\beta,\alpha}(t)$ which contains 
three index parameters $\it 0<\mu \leq 2$, $\it 0<\beta\leq 1$ and $\it \alpha >0$ and one 
time scale parameter $\it \xi$ (of units $\it sec^{-\beta}$). 
In order to derive the generalized space-time fractional diffusion equation 
it is convenient to proceed in the Fourier-Laplace domain. 
The Fourier-Laplace transform of the $\it {\mathbf P}^{(\mu)}_{\beta,\alpha}(t)$ is then with Eq.
(\ref{greensfunctionfrac}) given by the Montroll-Weiss equation
\begin{align}
\label{PKS}
{\tilde P}^{(\mu)}_{\beta,\alpha}(k,s) &= {\tilde P}({\mathbf k},0) \frac{\left(1-{\tilde \chi}_{\beta,\alpha}(s)\right)}{s}\frac{1}{(1-{\tilde \chi}_{\beta,\alpha}(s)
\lambda^{(\mu)}(\mathbf{k}))} = {\tilde P}({\mathbf k},0)
 \frac{s^{-1}}{\left(1+\frac{{\tilde \chi}_{\beta,\alpha}(s)}{(1-{\tilde \chi}_{\beta,\alpha}(s))}\frac{\eta^{\frac{\mu}{2}}({\mathbf k})}{{\cal K}^{(\mu)}}\right)} \\
&\label{PKS2} \approx {\tilde P}({\mathbf k},0)
 \frac{s^{-1}}{\left(1+\frac{{\tilde \chi}_{\beta,\alpha}(s)}{(1-{\tilde \chi}_{\beta,\alpha}(s))}\frac{k^{\mu}}{{\cal K}^{(\mu)}}\right)} ,
 \hspace{1cm} k\rightarrow 0  
\end{align}
containing also the Fourier transform $\it {\tilde P}({\mathbf k},0)$ of the initial condition (\ref{initialcondition}).
The exact equation (\ref{PKS}) can be rewritten as
\beq
\label{generalized-frac-diffeq-network}
-\frac{\xi^{\alpha}\eta^{\frac{\mu}{2}}({\mathbf k})}{{\cal K}^{(\mu)}}{\tilde P}^{(\mu)}_{\beta,\alpha}(k,s) =
(s^{\beta} +\xi)^{\alpha}{\tilde P}^{(\mu)}_{\beta,\alpha}(k,s) -\xi^{\alpha} 
{\tilde P}^{(\mu)}_{\beta,\alpha}(k,s) + \frac{\xi^{\alpha}-(s^{\beta}+\xi)^{\alpha}}{s} {\tilde P}({\mathbf k},0) ,\hspace{0.5cm} k_j
\in [-\pi,\pi] .
\eeq
Transforming back this equation into the causal time domain and by using 
Eqs. (\ref{initialcondition}) and (\ref{greensfunctionfrac})
yields the generalized time-fractional matrix equation
\beq
 \label{genfracdiffusionequationfractgen}
 -\frac{\xi^{\alpha}}{{\cal K}^{(\mu)}} {\mathbf L}^{\frac{\mu}{2}}\cdot {\mathbf P}^{(\mu)}_{(\beta,\alpha)}(t) = 
 _0\!\mathcal{D}_t^{\beta,\alpha} \cdot {\mathbf P}^{(\mu)}_{\beta,\alpha}(t) - 
 \xi^{\alpha} {\mathbf P}^{(\mu)}_{\beta,\alpha}(t) + 
 {\mathbf P}_0\left\{ \xi^{\alpha}\Theta(t) - K^{(0)}_{\beta,\alpha}(t)\right\} , \hspace{0.5cm} t \geq 0
\eeq
which we refer to as `generalized space-time fractional Kolmogorov-Feller equation' 
where $\it 0< \mu \leq 2$ with $\it 0<\beta \leq 1$ and $\it \alpha >0$. 
This equation was obtained and analyzed recently for normal walks ($\it \mu=2$) subordinated to a GFPP 
\cite{MichelRiascos2019,MichelitschRiascosGFPP2019}. Equations of the type (\ref{genfracdiffusionequationfractgen}) generally describe the
generalized space-time fractional diffusion on undirected networks connecting the network topology (contained in Laplacian matrix $\mathbf{L}$) 
with the GFPP-governed stochastic motion on the network.
We used notation $\it {\mathbf L}^{\frac{\mu}{2}}$ 
which denotes the fractional power of Laplacian matrix 
$\it {\mathbf L}$, and $\it {\mathbf P}^{(\mu)}_{(\beta,\alpha)}(t)$ the 
transition matrix with the initial condition $\it {\mathbf P}^{(\mu)}_{(\beta,\alpha)}(t=0)={\mathbf P}_0$ where all these matrices are 
defined in $ \it \mathbb{Z}^d$.
Since the $\it \mathbb{Z}^d$ is an infinite network, these are symmetric and circulant 
$\it \infty \times \infty$ matrices
with elements $\it L^{(\mu)}(\mathbf{p}-\mathbf{q}), P^{(\mu)}_{(\beta,\alpha)}(\mathbf{p}-\mathbf{q},t), 
{\mathbf P}_0(\mathbf{p}-\mathbf{q})$, respectively, where $\it \mathbf{p},\mathbf{q} \in \mathbb{Z}^d$.
In Eq. (\ref{genfracdiffusionequationfractgen}) we have introduced the causal convolution operator
$\it _0\!\mathcal{D}_t^{\beta,\alpha}$ and the causal function $\it K^{(0)}_{\beta,\alpha}(t)$
which were obtained in explicit forms \cite{MichelRiascos2019,MichelitschRiascosGFPP2019}
\beq
 \label{en-res-mittag-leffl-gn-kernel}
 \begin{array}{l}
 \it \ds {\cal D}^{\beta,\alpha}(t) = {\cal L}^{-1} \left\{(s^{\beta}+\xi)^{\alpha}\right\} = \frac{d^{\ceil{\alpha\beta}}}{dt^{\ceil{\alpha\beta}}} \left(\Theta(t) d^{\beta,\alpha}(t)\right)  \\ \\ \ds 
 \hspace{0.5cm} = \frac{d^{\ceil{\alpha\beta}}}{dt^{\ceil{\alpha\beta}}} (\Theta(t)d^{\beta,\alpha}(t-\tau))  = \left\{\begin{array}{l}
  \frac{d^{\ceil{\alpha\beta}}}{dt^{\ceil{\alpha\beta}}} 
  \left( \Theta(t) t^{\ceil{\alpha\beta}-\beta\alpha-1} 
  \sum_{m=0}^{\infty}\frac{\alpha!}{(\alpha-m)!m!} 
 \frac{(\xi  t^{\beta})^m}{\Gamma(\beta m +\ceil{\alpha\beta}-\beta\alpha)} \right),
 \hspace{0.5cm}\alpha\beta \notin \mathbb{N} \\ \\ \it \ds
 \frac{d^{\alpha\beta}}{dt^{\alpha\beta}}\left(
 \delta(t)+  \Theta(t) \frac{d}{dt}\sum_{m=0}^{\infty}\frac{\alpha !}{(\alpha-m)!m!} 
 \frac{(\xi t^{\beta})^m}{\Gamma(m\beta+1)} \right)  ,
\hspace{1cm} \alpha\beta \in \mathbb{N} \end{array}\right.
 \\ \\  \it \ds 
 \hspace{0.5cm} = \left\{\begin{array}{l} \frac{d^{\ceil{\alpha\beta}}}{dt^{\ceil{\alpha\beta}}} 
 \left( \Theta(t) t^{\ceil{\alpha\beta}-\beta\alpha-1}
 E_{\alpha,\beta,(\ceil{\alpha\beta}-\alpha\beta)}(\xi t^{\beta}) \right) ,\hspace{1cm} \alpha\beta \notin \mathbb{N}  \\ \\
\frac{d^{\alpha\beta}}{dt^{\alpha\beta}}\left(\delta(t) + 
\Theta(t)\frac{d}{dt}E_{\alpha,\beta,1}(\xi t^{\beta})\right) ,
\hspace{1cm} \alpha\beta \in \mathbb{N}. \end{array}\right. 
 \end{array}
\eeq
In these expressions we introduced the {\it ceiling function} $\it \ceil{\gamma}$ indicating the smallest integer
greater or equal to $\it \gamma$ and the function $E_{c,a,b}(z) = E^{-c}_{a,b}(-z)$ where $E^{u}_{v,w}(\zeta)$ indicating the
Prabhakar-Mittag-Leffler function (\ref{genmittag-Leff}). The operator $\it _0\!\mathcal{D}_t^{\beta,\alpha}$ acts on 
a causal distribution
$\it P(t)$ such as in Eq. (\ref{genfracdiffusionequationfractgen}) in the following way
\beq
\label{explicitf}
_0\!\mathcal{D}_t^{\beta,\alpha} \cdot P(t) = 
 \frac{d^{\ceil{\alpha\beta}}}{dt^{\ceil{\alpha\beta}}} \int_0^t d^{\beta,\alpha}(t-\tau)
  P(\tau) {\rm d}\tau .
\eeq
The function $\it K^{(0)}_{\beta,\alpha}(t)$ of equation (\ref{genfracdiffusionequationfractgen}) was obtained as
\beq
 \label{Kzeroexplicitform}
\it \ds  K^{(0)}_{\beta,\alpha}(t) ={\cal L}^{-1}\left\{\frac{(s^{\beta}+\xi)^{\alpha}}{s}\right\} = \left\{\begin{array}{l} \Theta(t)t^{-\alpha\beta} 
 E_{\alpha,\beta,1-\alpha\beta}(\xi t^{\beta}) ,\hspace{1cm} 0<\alpha\beta <1 \\ \\ \it \ds
 \frac{d^{\ceil{\alpha\beta}-1}}{dt^{\ceil{\alpha\beta}-1}}
   \left(\Theta(t) t^{\ceil{\alpha\beta}-\beta\alpha-1} 
 E_{\alpha,\beta,(\ceil{\alpha\beta}-\alpha\beta)}(\xi t^{\beta})\right)
 \hspace{0.5cm} \alpha\beta >1, \hspace{0.5cm} \alpha\beta \notin \mathbb{N} 
 \\ \\ \it \ds
 \frac{d^{\alpha\beta-1}}{dt^{\alpha\beta-1}}\left(\delta(t) + 
\Theta(t)\frac{d}{dt}E_{\alpha,\beta,1}(\xi t^{\beta})\right),
\hspace{1cm} \alpha\beta \geq 1 \in \mathbb{N}. 
                                    \end{array}\right. 
\eeq
Equation (\ref{genfracdiffusionequationfractgen}) governs the `microscopic' stochastic motions of the 
space fractional walk subordinated to a GFPP.
\section{\small DIFFUSION-LIMIT}
\label{diffusive-limit}

Our goal now is to determine the `diffusion-limit' of above stochastic motion to obtain a `macroscopic picture' on
spatial scales large compared to the lattice constant $\it 1$ of the $\it \mathbb{Z}^d$.
To this end it is sufficient to consider Montroll-Weiss equation (\ref{PKS}) for $\it k$ small. 
Then Eq. (\ref{generalized-frac-diffeq-network}) can be rewritten as
\beq
\label{Montroll-weiisdiffusive}
-\frac{\xi^{\alpha}k^{\mu}}{{\cal K}^{(\mu)}}{\tilde P}^{(\mu)}_{\beta,\alpha}({\mathbf k},s) \approx
(s^{\beta} +\xi)^{\alpha}{\tilde P}^{(\mu)}_{\beta,\alpha}({\mathbf k},s) -\xi^{\alpha} 
{\tilde P}^{(\mu)}_{\beta,\alpha}({\mathbf k},s) + \frac{\xi^{\alpha}-(s^{\beta}+\xi)^{\alpha}}{s} {\tilde P}({\mathbf k},0) ,\hspace{0.5cm} k\rightarrow 0 .
\eeq
In order to derive the `diffusive limit' which corresponds to the space-time representation of
this equation, it appears instructive to consider
the long-wave contribution of some kernels such as the fractional power of the Laplacian 
matrix in $\it \mathbb{Z}^d$. 
The fractional Laplacian matrix in $\it \mathbb{Z}^d$ has the canonic form 
\cite{TMM-APR-ISTE2019,TMM-APR-JPhys-A2017}
\beq
\label{left-hand}
[{\mathbf L}^{\frac{\mu}{2}}]_{{\mathbf p}-\mathbf{q}} = \frac{1}{(2\pi)^d} \int_{-\pi}^{\pi} {\rm d}k_1 \dots \int_{-\pi}^{\pi}{\rm d}k_d 
e^{i{\mathbf k}\cdot (\mathbf{p}-\mathbf{q})} \eta^{\frac{\mu}{2}}({\mathbf k}) ,\hspace{0.5cm} 0<\mu \leq 2
\eeq
where $\it \eta({\bf k})$ are the eigenvalues of the Laplacian matrix in $\it \mathbb{Z}^d$ defined by Eq. (\ref{eigvals}). Let us consider the contribution
generated by small $\it k\rightarrow 0$ by integrating over a small $\it d$-cube $\it |k_j| \leq k_c \ll 1$ 
around the origin (corresponding to very large wave-lengths), namely
\beq
\label{farfield}
\begin{array}{l} \it \ds
  [{\mathbf L'}^{\frac{\mu}{2}}]_{{\mathbf p}-\mathbf{q}}  \approx
\frac{1}{(2\pi)^d} \int_{-k_c}^{k_c} {\rm d}k_1 \dots \int_{-k_c}^{k_c}{\rm d}k_d 
e^{i{\mathbf k}\cdot (\mathbf{p}-\mathbf{q})} \eta^{\frac{\mu}{2}}({\mathbf k}) \\ \\ \it \ds
= \frac{h^d}{(2\pi)^d} \int_{-\pi h^{-\frac{1}{2}} }^{\pi h^{-\frac{1}{2}} } {\rm d}{\bar k}_1 \dots 
\int_{- \pi h^{-\frac{1}{2}}}^{\pi h^{-\frac{1}{2}} }{\rm d}{\bar k}_d  e^{i{\bar {\mathbf k}}\cdot (\mathbf{p}-\mathbf{q}) h } 
\eta^{\frac{\mu}{2}}(h{\bar {\mathbf k}}) .
\end{array}
\eeq
In the second line we have introduced a new wave vector $\it {\bar {\mathbf k}} $ with $\it k_j={\bar k}_j h \leq k_c$ with 
$\it k_c(h)= \pi h^{\frac{1}{2}} \rightarrow 0$ small thus 
 $\it 0 \leq |{\bar k}_j| \leq \pi h^{-\frac{1}{2}} \rightarrow \infty$ (where any exponent $\it 0<\delta <1$ could be used with 
 $\it |{\bar k}_j| \leq \pi h^{-\delta} \rightarrow\infty$ and $\it h{\bar k}_j \leq \pi h^{1-\delta} \rightarrow 0$).
 In this way the integral (\ref{farfield}) (rescaled $\it h^{-d}$) over small $\it k$ becomes an integral
 over the complete infinite $\it {\bar {\mathbf k}}$-space where in this integration 
 $\it k= {\bar k}h$ is small. Hence
 $\it (\eta(h{\bar k}))^{\frac{\mu}{2}} \approx 
 h^{\mu}{\bar k}^{\mu} \leq \pi^{\mu} h^{\frac{\mu}{2}} \ll 1 $ remains valid 
 in the entire region of integration in (\ref{farfield})$_2$.
Introducing 
the rescaled quasi-continuous very slowly varying `macroscopic' coordinates 
$\it (\mathbf{p}-\mathbf{q})h = \mathbf{r}-\mathbf{r}' \in h \mathbb{Z}^d$ of the nodes
we arrive at\footnote{This picture corresponds to the introduction of a lattice constant $\it h \rightarrow 0$.}
\beq
\label{farfield-htozero-res}  [\mathbf{L'}^{\frac{\mu}{2}}]_{{\mathbf p}-\mathbf{q}}  \approx  
  h^{\mu+d} \frac{1}{(2\pi)^d} \int_{-\infty}^{\infty} 
 {\rm d}{\bar k}_1\dots \int_{-\infty}^{\infty} {\rm d}{\bar k}_d  e^{i{\mathbf {\bar k}}\cdot (\mathbf{r}-\mathbf{r}')} {\bar k}^{\mu} = 
 h^{\mu+d} (-\Delta)^{\frac{\mu}{2}} \delta^d({\bf r}-{\bf r}') .
\eeq  
The new macroscopic coordinates $\it {\mathbf r} =h{\mathbf p} , \it {\mathbf r}'=h{\mathbf q} \in  \mathbb{R}^d $  are 
non-zero only for very large values of the integer values $\it p_j \sim h^{-1} \gg  1 $, $\it q_i h^{-1} \gg  1 $, 
i.e. the representation (\ref{farfield-htozero-res})
captures the far-field contribution $\it |\mathbf{p}-\mathbf{q}| \gg 1$. 
In Eq. (\ref{farfield-htozero-res}) $\it \Delta = \sum_{j=1}^d\frac{\partial^2}{\partial x_j^2}$ 
denotes the standard 
Laplacian with respect to the macroscopic coordinates $\it {\mathbf r}$.
The Fourier integral coincides up to the sign with
the kernel of the Riesz fractional derivative 
$\it -(-\Delta)^{\frac{\mu}{2}}\delta^{d}({\mathbf r}-{\mathbf r}')$ of the $\it \mathbb{R}^d$ 
(which has the eigenvalues $\it -k^{-\mu}$ and also is 
referred to as fractional Laplacian recovering for $\it \mu=2$ 
the standard Laplacian $\it \Delta $) \cite{TMM-APR-ISTE2019}.
It follows that Eq. (\ref{Montroll-weiisdiffusive}) can be 
transformed into the spatial (long-wave-) representation by

\beq
\label{renor}
 P^{(\mu)}(\mathbf{p}-\mathbf{q},t) \approx h^{d} {\bar P}^{(\mu)}(h(\mathbf{p}-\mathbf{q}),t) ,\hspace{1cm} h \rightarrow 0
\eeq
where we denote 
$\it  h(\mathbf{p}-\mathbf{q}) = {\mathbf r}-{\mathbf r}' \in \mathbb{Z}^d h \rightarrow \mathbb{R}^d$. The smooth field $\it {\bar P}^{(\mu)}({\bf r}-{\bf r}',t)$ 
introduced in asymptotic relation (\ref{renor})
indicates the macroscopic transition probability {\it density} kernel having physical units $\it cm^{-d}$.
By using Eqs. (\ref{greensfunctionfrac})-(\ref{PKS}) and 
$\it \lambda^{(\mu)}(h{\bar k})\approx 1-\frac{h^{\mu}{\bar k}^{\mu}}{{\cal K}^{(\mu)}}$ we arrive at
\beq
 \label{greensfunctionfrac-farfield}
  {\bar P}^{(\mu)}({ \mathbf r}-{\mathbf r}',t) \approx {\cal L}^{-1} \left\{
  \frac{s^{-1}}{(2\pi)^d} 
 \int_{-\infty}^{\infty}{\rm d}k_1\ldots \int_{-\infty}^{\infty}{\rm d}k_d e^{i{\mathbf {\bar k}}\cdot (\mathbf{r}-\mathbf{r}')} 
 \frac{{\bar P}(h{\bar k},0)}{\left(1+\frac{{\tilde \chi}_{\beta,\alpha}(s)}{1-{\tilde \chi}_{\beta,\alpha}(s)}\frac{h^{\mu}{\bar k}^{\mu}}{{\cal K}^{(\mu)}}\right)} \right\} 
\eeq
where the integration limits here can be thought to be generated by a limiting process 
$ \it \pm \lim_{h\rightarrow 0} \pi h^{-\frac{1}{2}} \rightarrow \pm \infty$ in the same way as in integral (\ref{farfield})
thus only small $\it h{\bar k} \leq \pi h^{\frac{1}{2}}$ in 
the integrand of (\ref{greensfunctionfrac-farfield}) are relevant.
Then let us rewrite Eq. (\ref{Montroll-weiisdiffusive}) in the Fourier-Laplace domain in the form
\beq
\label{Montroll-weiisdiffusive-rewri}
-\frac{h^{\mu}{\bar k}^{\mu}}{{\cal K}^{(\mu)}}{\tilde P}^{(\mu)}_{\beta,\alpha}(h{\bar k},s) \approx
\left[\left(1+\frac{s^{\beta}}{\xi}\right)^{\alpha} -1\right]
{\tilde P}^{(\mu)}_{\beta,\alpha}(h{\bar k},s) + \frac{{\tilde P}(h{\bar k},0)}{s} \left[1-\left(1+ \frac{s^{\beta}}{\xi}\right)^{\alpha}\right] ,\hspace{0.5cm} h\rightarrow 0 .
\eeq
We observe that $\it h\rightarrow 0$ makes left-hand side converging to zero 
(as within the integration limits of integral 
(\ref{farfield}) $\it k^{\mu}=h^{\mu}{\bar k}^{\mu} \leq \pi^{\mu} h^{\mu/2} \rightarrow 0$, i.e. 
only small $\it k=h{\bar k}$ are captured). 
In order to maintain the equality requires on the right-hand 
side of Eq. (\ref{Montroll-weiisdiffusive-rewri}) $\it \xi \rightarrow \infty$ thus 
we can expand $\it (1+ \frac{s^{\beta}}{\xi})^{\alpha} \approx 1+\frac{\alpha}{\xi}s^{\beta}$ and obtain
\beq
\label{diffusive-limitA}
-\frac{\xi h^{\mu}{\bar k}^{\mu}}{\alpha {\cal K}^{(\mu)}} {\tilde P}^{(\mu)}_{\beta,\alpha}(h{\bar k},s) 
\approx s^{\beta} {\tilde P}^{(\mu)}_{\beta,\alpha}(h{\bar k},s)
- {\tilde P}(h{\bar k},0)s^{\beta-1} .
\eeq
The existence of the diffusive limit requires the left-hand side of this equation 
to remain finite, i.e. $\it \xi h^{\mu} = const$ when $\it h\rightarrow 0$. 
It follows that $\it \xi$ then scales
as $\it \xi \sim h^{-\mu}$ leading to the new generalized diffusion constant 
\beq
\label{diffco}
{\cal A}=\frac{\xi h^{\mu}}{\alpha{\cal K}^{(\mu)}} > 0 ,\hspace{1cm} \alpha >0, \hspace{0.5cm} 0<\mu\leq 2 ,
 \eeq
having physical dimension $\it cm^{\mu}sec^{-\beta}$.
We obtain hence the {\it universal diffusion-limit} in the form of a 
{\it space-time fractional diffusion equation} of the form 
\beq
\label{universal-space-time-frac}
-{\cal A}\left(-\Delta\right)^{\frac{\mu}{2}} \cdot {\bar P}^{(\mu)}_{(\beta,\alpha)}(r,t) = 
_0 \!D_t^{\beta} \cdot {\bar P}^{(\mu)}_{\beta,\alpha}({\bf r},t) -  {\bar P}_0({\bf r}) \,\frac{t^{-\beta}}{\Gamma(1-\beta)}  \hspace{1cm} 0<\beta<1 .
\eeq
In this equation $\it _0  D_t^{\beta}\cdot (\dots) $ denotes the 
Riemann-Liouville fractional derivative of order $\beta$ 
(See Appendix \ref{causal}, Eq. (\ref{fracrieliou})), and $\it -\left(-\Delta\right)^{\frac{\mu}{2}}$ indicates
the Riesz-fractional derivative convolution operator (fractional Laplacian) 
in the $d$-dimensional 
infinite space\footnote{For explicit representations and evaluations, see e.g. \cite{TMM-APR-ISTE2019}.}.
The diffusion limit Eq. (\ref{universal-space-time-frac}) is
coinciding with 
a space-time fractional diffusion equation given by several authors 
\cite{SaichevZaslavski1997,MetzlerKlafter2004} in various contexts (among others). 
This equation is of the same type as the equation that occurs in the purely fractional 
Poisson process, i.e. for $\it \alpha=1$. We notice in view of the diffusion constant (\ref{diffco}) 
that index 
$\it \alpha$ appears in Eq. (\ref{universal-space-time-frac}) only 
as a scaling parameter.
The universal space-time fractional behavior of the diffusive 
limit reflects the {\it asymptotic universality} of the 
Mittag-Leffler waiting time PDF which was demonstrated by Gorenflo and Mainardi \cite{GorenfloMainardi2006}.
We emphasize the non-Markovian characteristics of this time fractional diffusion process in the range 
$\it 0<\beta<1$, i.e. when the waiting time PDF is fat-tailed. The non-markovianity is reflected by 
the occurrence of the slowly decaying
memory term $\it -{\bar P}_0({\bf r}) \,\frac{t^{-\beta}}{\Gamma(1-\beta)}$ in Eq. (\ref{universal-space-time-frac}) exhibiting a long-time memory
of the initial condition $\it {\bar P}^{\mu}_{\beta,\alpha}({\bf r},t=0) = {\bar P}_0({\bf r}) $.

Let us briefly consider the case when the walker at $t=0$ is in the origin. The initial condition then
is given by $ \it {\bar P}_0({\bf r}) = \delta^{d}({\bf r})$ 
and from Eqs. (\ref{diffusive-limitA})-(\ref{universal-space-time-frac}) follows that
\beq
\label{FourierLaplacetransmat}
{\tilde P}^{(\mu)}_{\beta,\alpha}(h{\bar k},s) = \frac{s^{\beta-1}}{{\cal A}{\bar k}^{\mu}+s^{\beta}} .
\eeq
In view of Eq. (\ref{fracPoissonsurv}) we obtain for the causal Fourier-time domain the solution 
\beq
\label{causaltimeFourier}
{\hat P}^{(\mu)}_{\beta,\alpha}(h{\bar k},t) = {\cal L}^{-1}\left\{\frac{s^{\beta-1}}{{\cal A}{\bar k}^{\mu}+s^{\beta}}\right\} = \Theta(t)
E_{\beta}(-{\cal A}{\bar k}^{\mu}t^{\beta}) ,\hspace{0.5cm} 0 < \beta \leq 1
\eeq 
where $\it E_{\beta}(z)$ denotes the Mittag-Leffler function defined in Eq. (\ref{mittag-le}) 
and for later convenience we included the Heaviside-step function 
$\it \Theta(t)$. In the space-time domain the transition probability kernel
is given by the Fourier inversion
\beq
\label{transprob-fourier}
{\bar P}^{(\mu)}_{\beta,\alpha}({\mathbf r},t) = \frac{\Theta(t)}{(2\pi)^d} \int_{-\infty}^{\infty}{\bar d}k_1 
\dots \int_{-\infty}^{\infty}{\bar d}k_d e^{i{\bar {\mathbf k}}\cdot{\mathbf r}}  E_{\beta}(-{\cal A}{\bar k}^{\mu}t^{\beta})
\eeq
where by accounting for $\it E_{\beta}(0)=1$ the initial condition 
$\it {\bar P}^{(\mu)}_{\beta,\alpha}({\mathbf r},0) = \delta^{d}({\mathbf r})$ is directly confirmed.
For $\it 0<\beta<1$ the Mittag-Leffler function
exhibits for $\it {\cal A}{\bar k}^{\mu}t^{\beta} \gg 1 $ inverse  power-law behavior, namely
\beq
\label{power-law}
{\hat P}^{(\mu)}_{\beta,\alpha}(h{\bar k},t) \approx {\cal L}^{-1}\left\{ {\cal A}^{-1} {\bar k}^{-\mu} s^{\beta-1}\right\} =  {\cal A}^{-1} {\bar k}^{-\mu} 
\,\frac{t^{-\beta}}{\Gamma(1-\beta)} .
\eeq
In the limit $\it \beta \rightarrow 1-0$ Eq. (\ref{universal-space-time-frac}) 
with $\it P_0({\bf r})= \delta^{d}({\mathbf r})$ takes for $\it 0<\mu<2 $ the form 
of a standard space-fractional L\'evy flight diffusion equation\footnote{See also Laplace transform 
(\ref{FourierLaplacetransmat}) for $\it \beta=1$ and Appendix \ref{causal}.}
\beq
\label{Levy-flight}
-{\cal A}\left(-\Delta\right)^{\frac{\mu}{2}} {\bar P}^{(\mu)}_{(1,\alpha)}(r,t) = \frac{\partial}{\partial t} {\bar P}^{(\mu)}_{(1,\alpha)}(r,t) 
- \delta^{d}({\mathbf r}) \delta(t) 
\eeq
where $\it 0<\mu \leq 2$ is admissible. This walk is Markovian and hence memoryless due to the immediate vanishing of the 
memory term $ \it - \delta^{(d)}({\mathbf r}) \delta(t)$ for $\it t>0$. For $\it \mu=2$ this equation recovers Fick's second 
law of normal diffusion.
For $\it \beta=1$ the Mittag-Leffler function in (\ref{causaltimeFourier}) turns into the form of
exponential $\it  E_1(-{\cal A}{\bar k}^{\mu}t) =  e^ {-{\cal A}{\bar k}^{\mu}t}$ 
thus the Fourier integral (\ref{transprob-fourier}) becomes
\beq
\label{gaussian-fourier}
{\bar P}^{(\mu)}_{(1,\alpha)}(r,t) =  \frac{\Theta(t)}{(2\pi)^d} \int_{-\infty}^{\infty}{\bar d}k_1 
\dots \int_{-\infty}^{\infty}{\bar d}k_d e^{i{\bar {\mathbf k}}\cdot{\mathbf r}}  e^{-{\cal A}{\bar k}^{\mu}t} , \hspace{0.5cm} 0<\mu \leq 2.
\eeq
One directly confirms that (\ref{gaussian-fourier}) solves (\ref{Levy-flight})\footnote{Where we 
take into account with $\it \delta(t)=\frac{d}{dt}\Theta(t)$ that $\it \frac{d}{dt}\left(\Theta(t)f(t)\right) = 
\delta(t)f(0)+ \Theta(t)\frac{d}{dt}f(t)$, and further properties are 
outlined in the Appendix \ref{causal}.} and is indeed the well-known expression for a symmetric L\'evy distribution in $\mathbb{R}^d$
\cite{MetzlerKlafter2000,MetzlerKlafter2004,TMM-APR-ISTE2019} (and many others). Further mention worthy is the case 
$\it \beta=1$ and $\it \mu=2$ for which
Eq. (\ref{Levy-flight}) recovers the form of a normal diffusion equation 
(Fick's second law) where (\ref{gaussian-fourier}) turns into the Gaussian distribution 
\beq 
\label{gaussian}
{\bar P}^{(2)}_{(1,\alpha)}(r,t) = \Theta(t)  \frac{e^{-\frac{r^2}{4{\cal A} t}}}{(4\pi {\cal A} t)^{\frac{d}{2}}}
\eeq
which indeed is the well-known causal solution of Fick's second law in $\it \mathbb{R}^d$.

\section{\small CONCLUSIONS}
\label{summary}
We developed a Montroll-Weiss CTRW model for space-fractional walks subordinated to a generalization of 
Laskin's fractional Poisson process, i.e. the fractional (long-range) jumps are
performed with waiting time PDF according to a `generalized fractional Poisson process' (GFPP).
We obtained a space-time fractional diffusion equation by defining a `well-scaled' diffusion limit
in the infinite $\it d$-dimensional integer lattice $\it \mathbb{Z}^d$ with a 
combined rescaling of space- and time-scales. The index $\it \alpha>0$ of the GFPP appears in this 
space-time fractional diffusion equation only as a scaling parameter. This diffusion equation
is of the same type as for $\it \alpha=1$ when the GFPP coincides with the pure Laskin's 
fractional Poisson process and exhibits for $\it 0<\beta<1$ non-Markovian features (long-time memory),
and for $\it \beta=1$ becomes a Markovian memoryless (L\'evy-flight) diffusion equation of standard Poisson.

The GFPP
contains three parameters, two index parameters $\it 0< \beta \leq 1$ and $\it \alpha >1$ 
and parameter $\it \xi$ 
defining a time scale. In the admissible range $\it 0<\beta\leq 1$, the waiting time PDF and the survival probability maintain their good properties
of complete monotony. For $\it \alpha=1$ and $\it 0<\beta<1$ the equations of Laskin's fractional Poisson process 
and for $\it \alpha=1$, $\it \beta=1$ the classical equations of the standard Poisson process are recovered, respectively.
Some of the discussed results were obtained in recent papers \cite{MichelRiascos2019,MichelitschRiascosGFPP2019}.
Generalizations of fractional diffusion as analyzed in the present paper 
are interesting models for a better understanding of the stochastic dynamics in complex systems. 
Since these models offer more
parameters they are susceptible to be adopted to describe real-world situations.

\begin{appendix}

\section{APPENDIX}
\subsection{\small LAPLACE TRANSFORMS AND FRACTIONAL OPERATORS}
\label{causal}
Here we derive briefly some basic mathematical apparatus used in the paper 
in the context of causal functions and distributions involving fractional operators and Heaviside-calculus. All functions and distributions 
considered are to be conceived as generalized functions and distributions in the Gelfand-Shilov sense \cite{GelfangShilov1968}.
Let us first introduce the Heaviside step-function
\beq
 \label{Heavisidestep}
 \Theta(t) = \left\{\begin{array}{l}  1 ,\hspace{1cm} t \geq 0 \\ \\
                     0 ,\hspace{1cm} t < 0.
                    \end{array}\right. 
\eeq
A function is causal if it has the form $\it \Theta(t) f(t)$, i.e. is null for $\it t<0$ and non-vanishing only 
for non-negative times $\it t$. 
We introduce the Laplace transform of $\it \Theta(t) f(t)$ by
\beq
 \label{Laplace-trans}
 {\tilde f}(s) = {\cal L}(f(t)) =  \int_{-\infty}^{\infty} e^{-st} \Theta(t)f(t){\rm d}t = \int_{0}^{\infty}f(t)e^{-st}{\rm d}t ,\hspace{1cm} s=\sigma+i\omega
\eeq
with suitably chosen $\it \sigma >\sigma_0$ in order to guarantee convergence of (\ref{Laplace-trans}). In view of the fact that (\ref{Laplace-trans}) can be read as Fourier transform of the causal function $e^{-\sigma t}f(t)\Theta(t)$
it is straightforward to see that the Laplace inversion corresponds to the representation of this function as Fourier integral namely
\beq
 \label{Fourierint}
 e^{-\sigma t}f(t)\Theta(t) = \frac{1}{2\pi} \int_{-\infty}^{\infty} e^{i\omega t} {\tilde f}(\sigma+i\omega){\rm d}\omega
\eeq
which can be rewritten as
\beq 
 \label{Laplacebachtrafo}
 f(t)\Theta(t) = \frac{ e^{\sigma t} }{2\pi} \int_{-\infty}^{\infty} e^{i\omega t} {\tilde f}(\sigma+i\omega){\rm d}\omega = \frac{1}{2\pi i} \int_{-i\infty}^{+i\infty} e^{st} {\tilde f}(s){\rm d}s .
\eeq
Sometimes when there is no time derivative involved we skip the Heaviside $\it \Theta(t)$-function implying that all expressions are written for $t\geq 0$.
Then we mention that
\beq
 \label{trivial}
 \Theta(t)f(t) = \int_{-\infty}^{\infty} \delta(t-\tau)\Theta(\tau)f(\tau){\rm d}\tau 
\eeq
and introduce the shift operator $\it e^{-\tau \frac{d}{dt}}$ acting on a function $\it g(t)$ as $\it e^{-\tau \frac{d}{dt}}g(t)=g(t-\tau)$ thus
\beq
\label{shift-op}
 e^{-\tau \frac{d}{dt}} \delta(t) = \delta(t-\tau) .
\eeq
Substituting this relation into (\ref{trivial}) yields 
\beq
 \label{non-trivial}
 \Theta(t)f(t) =  \left\{\int_{-\infty}^{\infty} e^{-\tau\frac{d}{dt}} \Theta(\tau)f(\tau){\rm d}\tau\right\} \,\, \delta(t) ={\cal L}^{-1}\{ {\tilde f}(s) \} = {\tilde f}\left(\frac{d}{dt}\right) \,\, \delta(t)
\eeq
where the operator $\it {\tilde f}(\frac{d}{dt})$ is related with the Laplace transform (\ref{Laplace-trans}) by replacing $\it  s \rightarrow \frac{d}{dt}$.
Eq. (\ref{non-trivial}) is the operator representation of the causal function $\Theta(t)f(t)$. A convolution of two causal functions $\it \Theta(t)f(t), g(t)\Theta(t)$ then can be represented by
\beq
 \label{convolutions}
 \begin{array}{l}
 \ds \int_{0}^t g(t-\tau)f(\tau){\rm d}\tau =  \int_{-\infty}^{\infty} \int_{-\infty}^{\infty} \delta(t-\tau_1-\tau_2) g(\tau_1)\Theta(\tau_1)f(\tau_2)\Theta(\tau_2) {\rm d}\tau_1{\rm d}\tau_2 ,\hspace{1cm} t>0  \\ \\
 \ds \left(\int_{-\infty}^{\infty} e^{-\tau_1\frac{d}{dt}} \Theta(\tau_1)f(\tau){\rm d}\tau_1\right)  
 \left(\int_{-\infty}^{\infty} e^{-\tau_2\frac{d}{dt}} \Theta(\tau_2)f(\tau){\rm d}\tau_2\right) \delta(t) \\ \\
 \ds ={\cal L}^{-1}\{{\tilde g}(s){\tilde f}(s)\} 
  = {\tilde f}\left(\frac{d}{dt}\right) {\tilde g}\left(\frac{d}{dt}\right) \,\, \delta(t) = {\tilde g}\left(\frac{d}{dt}\right) {\tilde f}\left(\frac{d}{dt}\right) \,\ \delta(t)
 \end{array}
\eeq
where it has been used $\it \delta(t-\tau_1-\tau_2) =  e^{-(\tau_1+\tau_2)\frac{d}{dt}}\delta(t)$. We observe that in 
(\ref{non-trivial}) and (\ref{convolutions}) the Laplace variable is replaced $\it s \rightarrow  \frac{d}{dt}$ 
in the causal time domain.
By considering $\it {\bar f}(t)\Theta(t)= f(t)\Theta(t) e^{-\lambda t}$ we observe that
\beq
 \label{consderexp}
 f(t)\Theta(t) e^{-\lambda t} = \left\{\int_{-\infty}^{\infty} e^{-\tau(\lambda +\frac{d}{dt})} f(\tau)\Theta(\tau) {\rm d}\tau\right\} \delta(t) = {\tilde f}\left(\lambda+\frac{d}{dt}\right) \delta(t)
\eeq
where $\it {\tilde f}(s)= {\cal L}\{f(t)\}$. We are especially dealing with normalized (probability-) distributions 

\beq
\label{norma}
{\tilde f}(s=0)= 1 = \int_{-\infty}^{\infty}\Theta(t) f(t){\rm d}t.
\eeq
A very important consequence of relations (\ref{non-trivial}) and (\ref{convolutions}) is that they can be used to solve
differential equations and to determine causal Green's functions. As a simple example consider the trivial algebraic equation in the Laplace domain
\beq
\label{simplerel}
(s+\xi) \,\, \frac{\xi}{(s+\xi)} = \xi  , \hspace{1cm} \xi >0
\eeq
takes with $\it {\cal L}^{-1}\left\{s+\xi\right\} = \frac{d}{dt}+\xi$ and 
$\it {\cal L}^{-1}\left\{\frac{\xi}{s+\xi}\right\} = \Theta(t) \xi e^{-\xi t}$ 
where on the right-hand side is used that
$\it {\cal L}^{-1}\{1\}  = \delta(t) $. 
In the causal time domain (\ref{simplerel}) then gives
the representation
\beq
\label{saulastimeeq}
\left(\frac{d}{dt}+\xi\right)\, \left(  \xi \Theta(t) e^{-\xi t} \right) =  \xi \delta(t) ,
\eeq
result which is straightforwardly confirmed, i.e. the normalized causal Green's function of $\it \frac{d}{dt}+\xi$ is directly obtained as
$\it (\frac{d}{dt}+\xi)^{-1} \xi \delta(t) = \Theta(t) \xi e^{-\xi t}$ where it is important 
that the  $\it \Theta(t)$-function is taken into account in the Laplace inversion.

A less trivial example is obtained when considering fractional powers of operators. For instance let us consider in the Laplace domain the
equation
\beq
\label{simplerelfract}
(s^{\beta}+\xi) \, \frac{\xi}{(s^{\beta}+\xi)} = \xi  , \hspace{1cm} \xi >0, \hspace{1cm} 0< \beta \leq 1
\eeq
which writes in the time domain
\beq 
\label{timedomainefracteq}
\left\{\left(\frac{d}{dt}\right)^{\beta}+\xi\right\} \, \Theta(t) g_{\beta,\xi}(t) = \xi \delta(t)
\eeq
where the fractional derivative $\left(\frac{d}{dt}\right)^{\beta}$ is determined subsequently. The causal
Green's function $\it \Theta(t) g_{\beta,\xi}(t)$ is obtained from the Laplace inversion  

\beq
\label{fraclaplaceinv}
g_{\beta,\xi}(t)={\cal L}^{-1}\left\{\frac{\xi}{s^{\beta}+\xi}\right\}.
\eeq
The inversion is performed directly when taking into account
\beq
\label{takeintoaccount}
\frac{\xi}{(s^{\beta}+\xi)}  =  s^{-\beta} \frac{\xi}{(1+\xi s^{-\beta})} = \xi
\sum_{n=0}^{\infty} (-1)^n \xi^n s^{-\beta(n+1) }, \hspace{1cm} \sigma=\Re\{s\} > \xi^{\frac{1}{\beta}}
\eeq
where $\it \sigma=\Re\{s\} > \xi^{\frac{1}{\beta}}$ guarantees convergence of this geometric series $\it \forall \omega = \Im\{s\}$, 
i.e. for the entire
interval of integration of the corresponding Laplace inversion integral (\ref{consderexp}).
On the other hand we have
\beq 
\label{hrelation}
s^{-\mu} = {\cal L}\left\{ \Theta(t)\frac{t^{\mu-1}}{\Gamma(\mu)} \right\}  ,\hspace{1cm}  \mu >0 , \hspace{1cm} \sigma > 0 ,
\eeq 
where we use the notation $\it \Gamma(\xi+1)= \xi ! $ for the Gamma-function.
We then arrive at
\beq
\label{Mittag-Leffler-density-1}
\begin{array}{l} \it \ds 
\Theta(t) g_{\beta,\xi}(t)={\cal L}^{-1}\left\{\frac{\xi}{\xi+s^{\beta}}\right\}  \\ \\ \it \ds \hspace{0.5cm} 
=  \sum_{n=0}^{\infty} (-1)^n \xi^{n+1} {\cal L} ^{-1}\left\{s^{-\beta(n+1) }\right\}=
\Theta(t) \sum_{n=0}^{\infty} (-1)^n \xi^{n+1} 
\frac{t^{n\beta +\beta-1}}{\Gamma(n\beta +\beta)}
\end{array}
\eeq
with
\beq
\label{Mittag-Leffler-density-result}
 g_{\beta,\xi}(t) = \xi t^{\beta-1} \sum_{n=0}^{\infty}\frac{(-\xi t^{\beta})^n}{\Gamma(n\beta+\beta)} =  \xi t^{\beta-1} E_{\beta,\beta}(-\xi t^{\beta}) .
\eeq
Here we have introduced the generalized Mittag-Leffler function, 
e.g. \cite{Gorenflo2007,HauboldMathaiSaxena2011,GorenfloKilbasMainardiRogosin2014}
\beq
\label{genmiLeff}
E_{\beta,\gamma}(z) =  \sum_{n=0}^{\infty}  \frac{z^n}{\Gamma(\beta n+\gamma)} ,\hspace{1cm} \beta, \gamma >0 ,\hspace{0.5cm} z \in \mathbb{C}
\eeq
It follows that $\it \xi$ is a dimensional constant having units $\it sec^{-\beta}$ so that 
(\ref{Mittag-Leffler-density-result}) has physical dimension of
$\it sec^{-1}$ of a density. The result (\ref{Mittag-Leffler-density-result}) also is referred to as 
{\it Mittag-Leffler density} and represents the waiting time density of Eq. (\ref{Mittag-Leffler-density})
of the fractional Poisson renewal process introduced by Laskin \cite{Laskin2003}.
Generally Mittag-Leffler type functions play a major role in time fractional dynamics.
We further often use the Mittag-Leffler function which is defined as, e.g. 
\cite{Gorenflo2007,HauboldMathaiSaxena2011,GorenfloKilbasMainardiRogosin2014}
\beq 
 \label{mittag-le} 
 E_{\beta}(z) = \sum_{n=0}^{\infty} \frac{z^n}{\Gamma(\beta n+1)} ,\hspace{1cm} \beta >0 ,\hspace{0.5cm} z \in \mathbb{C}
\eeq
where with (\ref{genmiLeff}) we have $\it E_{\beta}(z)=E_{\beta,1}(z)$.
The Mittag-Leffler function has the important property that for $\it \beta=1$ it recovers the exponential
$\it E_{1}(z)= e^{z}$.
\\[2ex]

\noindent {\it Riemann-Liouville fractional integral and derivative}
\\ \\
Now let us derive the kernel of the fractional power of time-derivative operator of Eq. (\ref{timedomainefracteq})
where we consider now exponents $\it \gamma>0$.
This kernel is then obtained with above introduced methods in the following short way
\beq
 \label{part1}
 \begin{array}{l}
\it \ds {\cal L}^{-1}\{s^{\gamma}\}=  {\cal L}^{-1} \{s^{\ceil{\gamma}} s^{\gamma -\ceil{\gamma}}\} = 
  e^{\sigma t}\left(\sigma+\frac{d}{dt}\right)^{\ceil{\gamma}}\left(\sigma+\frac{d}{dt}\right)^{\gamma-\ceil{\gamma}} \delta(t) 
  ,\hspace{1cm} \gamma >0 , \gamma \notin \mathbb{N}
  \\ \\ \it \ds
 \hspace{0.5cm}  =
  e^{\sigma t}\left(\sigma+\frac{d}{dt}\right)^{\ceil{\gamma}}
 \int_{-\infty}^{\infty}\frac{{\rm d}\omega}{(2\pi)}e^{i\omega t}
 (\sigma+i\omega)^{\gamma-\ceil{\gamma}}  \\ \\ \it \ds \hspace{0.5cm} =
 e^{\sigma t}\left(\sigma+\frac{d}{dt}\right)^{\ceil{\gamma}} \left\{ e^{-\sigma t}\Theta(t)  
 \frac{t^{\ceil{\gamma}-\gamma-1}}{(\ceil{\gamma}-\gamma -1)!}\right\} = 
 \frac{d^{\ceil{\gamma}}}{dt^{\ceil{\gamma}}} \left(\Theta(t) 
 \frac{t^{\ceil{\gamma}-\gamma-1}}{\Gamma(\ceil{\gamma}-\gamma)} \right) .
 \end{array}
\eeq
Here we introduced the {\it ceiling function} $\it \ceil{\gamma}$ indicating the smallest integer
greater or equal to $\it \gamma$. In this way the Fourier integral in the second line is 
integrable around $\it \omega=0$ $\it \forall \sigma \geq 0$ since
$\it \gamma -\ceil{\gamma} >-1 $.
We then obtain for the Laplace inversion 
\beq
 \label{Lplaceinversionnegativegamma}
 {\cal L}^{-1}\{s^{-\gamma}\}= \Theta(t) \frac{t^{\gamma-1}}{\Gamma(\gamma)} ,\hspace{1cm}  \gamma >0
\eeq
as a fractional generalization of integration operator. This kernel indeed can be identified 
with the kernel of the {\it Riemann-Liouville fractional 
integral} operator of order $\it \gamma$ \cite{OldhamSpanier1974,MillerRoss1993,SamkoKilbasMarichev1993} 
which recovers for $\gamma \in \mathbb{N}$ the
multiple integer order integrations.  

On the other hand the kernel (\ref{part1}) with explicit representation in (\ref{part1})$_3$ can 
be conceived as the `fractional derivative' operator $\it (\frac{d}{dt})^{\gamma}$.
The fractional derivative acts on causal functions $\it \Theta(t)f(t)$ as 
\beq
 \label{fractderiv}
 \begin{array}{l}
 \it \ds {\cal L}^{-1}\{s^{\gamma}\} \cdot f(t)\Theta(t) =: _0\!D_t^{\gamma} f(t) = \frac{d^{\ceil{\gamma}}}{dt^{\ceil{\gamma}}}\int_{-\infty}^{\infty}
 \left\{ \Theta(t-\tau) 
 \frac{(t-\tau)^{\ceil{\gamma}-\gamma-1}}{{\Gamma(\ceil{\gamma}-\gamma)}}  \right\} f(\tau)\Theta(t) {\rm d}\tau, \\ \\
\it \ds _0\!D_t^{\gamma} f(t) = \frac{1}{{\Gamma(\ceil{\gamma}-\gamma)}}  \frac{d^{\ceil{\gamma}}}{d\tau^{\ceil{\gamma}}} \int_0^t (t-\tau)^{\ceil{\gamma}-\gamma-1} f(\tau){\rm d}\tau .
 \end{array} \hspace{1cm} \gamma>0 
\eeq
We identify in the last line this operator with the {\it Riemann-Liouville fractional derivative} 
\cite{OldhamSpanier1974,MillerRoss1993,SamkoKilbasMarichev1993} 
which recovers for $\gamma \in \mathbb{N}$ integer-order standard derivatives.
We emphasize that (\ref{part1}) requires causality, i.e. a distribution 
of the form $\it f(t)\Theta(t)$ thus the Laplace transform captures the entire non-zero contributions 
of the causal distribution. 
In the diffusion equation (\ref{universal-space-time-frac}) 
the Riemann-Liouville fractional derivative is of order $\it 0<\beta<1$ with $\it \ceil{\beta}=1$. In this case 
(\ref{fractderiv}) then yields has representation
\beq
\label{fracrieliou}
\ds _0\!D_t^{\beta} f(t) =\frac{1}{\Gamma(1-\beta)}\frac{d}{dt}\int_0^t (t-\tau)^{-\beta}f(\tau){\rm d}\tau 
,\hspace{1cm} 0<\beta<1 , \hspace{0.5cm} t>0 .
\eeq

\end{appendix}

\end{document}